\documentclass[universe,article,accept,pdftex,moreauthors]{Definitions/mdpi}
\usepackage{bm}
\usepackage{siunitx}
\firstpage{1}
\pubvolume{1}
\issuenum{1}
\articlenumber{0}
\pubyear{2024}
\copyrightyear{2024}
\externaleditor{Academic Editor: Firstname \\Lastname}
\datereceived{11 July 2024}
\daterevised{11 August 2024} 
\dateaccepted{14 August 2024}
\datepublished{19 August 2024}
\datecorrected{} 
\dateretracted{} 
\hreflink{https://doi.org/} 
\doinum{}
\CorrStatement{yes}  

\Title{Quantum Effects on Cosmic Scales as an Alternative to Dark Matter and Dark Energy}

\TitleCitation{Quantum Effects on Cosmic Scales as an Alternative to Dark Matter and Dark Energy}

\Author{{Da-Ming Chen} 
 $^{1,2,}$* and Lin Wang $^{3}$}

\AuthorNames{Da-Ming Chen and Lin Wang}

\AuthorCitation{{Chen, D.-M.;} 
 Wang, L.}

\address{%
$^{1}$ \quad National Astronomical Observatories, Chinese Academy of Sciences, 20A Datun Road, Chaoyang District, Beijing 100101, China\\
$^{2}$ \quad School of Astronomy and Space Science, University of Chinese Academy of Sciences, Beijing 100049, China\\
$^{3}$ \quad School of Physics and Electronics, Henan University, Kaifeng 475004, China; {wl010@bao.ac.cn} 
}

\corres{Correspondence: cdm@nao.cas.cn}

\abstract{The spin-torsion theory is a gauge theory approach to gravity that expands upon Einstein's general relativity (GR) by incorporating the spin of microparticles. In this study, we further develop the spin-torsion theory to examine spherically symmetric and static gravitational systems that involve free-falling macroscopic particles. We posit that the quantum spin of macroscopic matter becomes noteworthy at cosmic scales. We further assume that the Dirac spinor and Dirac equation adequately capture all essential physical characteristics of the particles and their associated processes. A crucial aspect of our approach involves substituting the constant mass in the Dirac equation with a scale function, allowing us to establish a connection between quantum effects and the scale of gravitational systems. This mechanism ensures that the quantum effect of macroscopic matter is scale-dependent and diminishes locally, a phenomenon not observed in microparticles. For any given matter density distribution, our theory predicts an additional quantum term, the quantum potential energy (QPE), within the mass expression. The QPE induces time dilation and distance contraction, and thus mimics a gravitational well.  When applied to cosmology, our theory yields a static cosmological model. The QPE serves as a counterpart to the cosmological constant introduced by Einstein to balance gravity in his static cosmological model. The QPE also offers a plausible explanation for the origin of Hubble redshift (traditionally attributed to the universe's expansion). The predicted luminosity distance--redshift relation aligns remarkably well with SNe Ia data from the cosmological sample of SNe Ia.  In the context of galaxies, the QPE functions as the equivalent of dark matter. The predicted circular velocities align well with rotation curve data from the SPARC (Spitzer Photometry and Accurate Rotation Curves database) sample.  Importantly, our conclusions in this paper are reached through a conventional approach, with the sole assumption of the quantum effects of macroscopic matter at large scales, without the need for additional modifications or assumptions.}


\begin{document}

\section{Introduction}
The primary challenge facing the current theory of gravity, the~Einstein--Newton theory, lies in the unresolved mysteries of dark matter and dark energy. To~date, there has been a lack of direct observational evidence confirming the existence of these enigmatic components. Extensive efforts are being made to address this topic. On~the theoretical front, much attention is directed towards modifying or extending the Einstein--Newton theory of gravity to align with astronomical observations. For~instance, one can enhance the standard Lagrangian in general relativity by incorporating higher-order curvature corrections~\citep{lovelock1971einstein,Lovelock1972,PhysRevLett.55.2656,2005GReGr..37.1869K,Oikonomou_2021,brassel2022charged}, or~formulate non-linear Lagrangians~\citep{buchdahl1970non,goswami2014collapsing}. Other relevant examples include modified Newtonian dynamics (MOND)~\cite{1983ApJ...270..365M,famaey:hal-02927744} and its relativistic version~\cite{2004PhRvD..70h3509B}, as~well as conformal gravity~\cite{Mannheim_1997,MANNHEIM2006340}.

In this paper, we explore the concept of quantum spacetime at cosmic scales as a potential alternative to dark matter and dark energy. This research represents a significant extension of our previous study~\citep{Chen_2022}. Rather than relying on previous assumptions, we explicitly propose scale-dependent quantum properties of spacetime. This proposition is motivated by our interest in potentially moving away from the notion of a preferred absolute spacetime ({PAS}) for the universe, as~implied by general relativity (GR) and illustrated in the standard $\Lambda$CDM cosmology. {It is believed that the PAS is theoretically characterized by the Friedmann--Lemaître--Robertson--Walker  (FLRW)  metric and empirically supported by the cosmic microwave background (CMB). One aim of this  study is to  demonstrate  that  the  notion  of  a  PAS  is  not  inherently  valid  when  considering  the  quantum  effects  of  macroscopic  matter  at  large  scales. }

Before going further, it is imperative to clarify our actual motivation in order to prevent any possible misunderstandings. For~ease of reference, we categorize the universe into three distinct classes based on spatial scales: the cosmic scale (encompassing galaxies to the entire universe), the~macroscopic scale (ranging from everyday life to the size of the solar system), and~the microscopic scale (where quantum mechanics becomes significant, as is well-known). According to the mainstream view, the~quantum effect for objects at the macroscopic and cosmic scales can be safely disregarded, due to their large mass. This is known as the classical limit. {}{We will show}, however, that quantum effects are dependent on the scale of the gravitational system considered and that~the ``classical limit'' is only reached at macroscopic scale. In~other words, quantum effects are significant not only at microscopic scale but also at cosmic scale, and~can only be ignored at macroscopic scale. Intuitively, when we study cosmology, any observers who perceive distant galaxies in the universe, would expect them to exhibit quantum behavior resembling that of microscopic particles, due to their distant location. As~for galactic dynamics, ~quantum effects should also be taken into account for galaxies, although~in the central part, they can be ignored due to {the mechanism presented in this study.}

It  is  helpful  to  discuss  the  reasons  behind  generalizing  the  quantum  effects  observed  in  microscopic  particles  to  macroscopic  ones,  as~ well  as  why  the  strength  of  these  generalized  quantum  effects  appears  to  increase  with  the  spatial  scale,  based  on  our  previous  paper~\cite{Chen_2022}. It  is  believed  that  gravity  turns  disorder  into  order,  and~ order  is  fundamental  to  space,  time,  and~ spacetime---or,  as~ we  might  say,  spacetime  is  nothing  but  the  continuous  ordering  of  events.  Thus,  without~ gravity,  there  would  be  no  spacetime.  In~ any  local  inertial  frame,  gravity  is  still  present,  but~ the  net  gravitational  force  is  canceled  out  by  the  inertial  force,  leaving  order  or  flat  spacetime  behind. In~ special  relativity,  rigid  rods  are  used  to  create  coordinate  lattices,  ensuring  that  every  event  has  a  position  at  any  given  time.  According to  GR,  the~ global  structure  of  curved  spacetime  can  still  be  described  using  arbitrarily  curved  rigid  rods  (depending  on  the  matter  distribution)  of  arbitrary  length.  However,  quantum  mechanics  reveals  that  microscopic  particles  can  escape  the  order  depicted  by  spacetime  and  thus  elude  the  control  of  gravity.  The~ quantum  effects  of  electrons,  for~ instance,  can  be  attributed  to  their  quantum  randomness  (also  known  as  intrinsic  randomness),  which  is  markedly  different  from  classical  randomness  (or  apparent  randomness),  exemplified  by  Brownian  motion. A~ free  electron  deviates  from  a  straight  line  or  a  curved  geodesic,  which  implies  that  gravity  is  not  strong  enough  to  bring  it  into  order.  This  explains  quantum  effects  from  a  geometric  point  of  view.  Gravity  does  bring  macroscopic  matter  into  order  in  the  sense  that  any  free  macroscopic  object  will  move  along  a  straight  line  or  a  curved  geodesic.  This  fact  is  fundamental  to  both  Newtonian  theory  and  GR. However,  if~ we  assume  that  all  matter,  irrespective  of  its  mass,  has  two  opposite  properties---gravity  and  quantum  randomness---then  it  is  possible  that  gravity  is  not  strong  enough  to  bring  any  distant  macroscopic  object  into  order,  just  as occurs  with  microscopic  particles.  If~ our  assumption  is  true,  then  general  relativity  (GR)  is  valid  only  approximately  within  a  sufficiently  small  neighborhood  of  any  point  in  the  spacetime  manifold.  In~ this  case,  the~ uncertainty  in  matter  distribution,  originating  from  quantum  randomness,  will  accumulate  with  spatial  distance.  Or,  put  another  way,  the~ strength  of  quantum  effects  appears  to  increase  with  spatial  scale. On~the other hand, since the spin-induced torsion arising from macroscopic matter vanishes locally, the~geodesics of test particles are precisely defined. This  serves  as  the  physical  foundation  for  physicists  to  construct  coordinate  lattices.  As~ a  result,  in~ general,  spacetime  is  not  only  curved  but  also  flexible.  It  turns  out  that  the  preferred  absolute  spacetime  (PAS),  which  is  determined  by  all  matter  in  the  universe,  loses  its  meaning  when  scale-dependent  quantum  effects  come  into  play.  Any  new  fundamental  assumptions  about  physical  laws  should  be  falsifiable.  The~ rest  of  this  study  represents  a  first  attempt  along  this  line.    

We will illustrate the integration of our assumption with the Einstein equations. It is natural for us to generalize the Dirac theory to describe the {macroscopic} matter content {of gravitational systems with cosmic scale}. 

There are two approaches to interpreting quantum mechanics. The~first is the standard view, which asserts that microscopic particles cannot have continuous trajectories and that non-commuting observables (such as position and momentum) must satisfy the uncertainty principle. The~second approach is known as the causal interpretation, which posits that each microscopic particle has a continuous trajectory and replaces the uncertainty principle with the concept of quantum potential \citep{bohm1995undivided,holland1995quantum}. Both approaches yield equivalent predictions for experimental results. However, the~causal interpretation, particularly in its geometric algebra (GA) version, provides profound insights into the quantum nature of spacetime and matter~\citep{Hestenes_1979,1967JMP.....8..798H,1973JMP....14..893H,2003AmJPh..71..691H}. In~this paper, similarly to in our previous work, we utilize spacetime algebra (STA)~\citep{1966STA+Hestenes,2003AmJPh..71..691H,doran2003geometric} as our mathematical language. STA is constructed from a Minkowskian vector space and provides a straightforward geometric understanding of Dirac theory. It provides a natural link to classical mechanics, and~for expressing Dirac theory with observables, it offers enhanced computational efficiency and capabilities~\citep{1973JMP....14..893H} when compared to the tensor analysis method ~\citep{Takabayasi1957}. The~fundamentals of STA  are outlined in Appendix~\ref{sec:STA}. In~the STA version of Dirac theory, known as real Dirac theory, Hestenes identified the complex number present in the matrix version of Dirac theory as the spin plane~\citep{1967JMP.....8..798H}. Furthermore, since the complex number is always accompanied by the Planck constant $\hslash$, a~rigorous derivation of the Schrodinger theory from the Pauli or Dirac theory implies that the Schrodinger equation describes an electron in an eigenstate of spin, rather than, as~commonly believed, an~electron without spin~\citep{1971AmJPh..39.1013H}. Correspondingly, there are two gauge theories of gravity concerning our research work: one is the tensor analysis approach~\citep{1976RevModPhys.48.393}, and~the other is the gauge theory gravity (GTG) developed by the Cambridge group using STA~\citep{1998RSPTA.356..487L}. These~two theories are nearly equivalent, but~the latter is conceptually clearer and technically more powerful for comprehending and calculating the challenges encountered in quantum mechanics and gravity. Therefore, we choose to adopt GTG in this paper. The~fundamentals of GTG and its applications to spin-$\frac{1}{2}$ particles are summarized in Appendix~\ref{sec:GTG}.

It is now widely acknowledged that the quantum random motion of spin-$\frac{1}{2}$ particles can be fully described by their spin. In~the presence of gravity, the~spin gives rise to the torsion of spacetime~\citep{1976RevModPhys.48.393,1998RSPTA.356..487L,1998JMP....39.3303D, doran2003geometric,2005FoPh...35..903H,sym16010067}. {In particular, the~effects of spin-torsion in GTG were investigated in Ref.~\cite{1998JMP....39.3303D}. GTG is nearly equivalent to Einstein--Cartan gravity~\cite{1976RevModPhys.48.393}. The~ stress-energy  tensor  derived  from  the  Dirac  theory  contains  an  asymmetric  component,  representing  the  contribution  of  quantum  spin.  As~ a  consequence,  the~ metric  of  spacetime  as  predicted  by  the  generalized  Einstein  equations  includes  a  component  that  accounts  for  the  torsion  of  spacetime.  As~will be shown later, the~torsion term appearing in the metric represents a modification to GR. To  date,  the~ problem  of  finding  solutions  for  a  Dirac  field  coupled  to  gravity  in  a  self-consistent  manner  has primarily  been  considered   for  microscopic  particles.  To~ our  knowledge,  there  is  only  one  work  that  explored the massive, non-ghost cosmological solutions for the Dirac field coupled self-consistently to gravity~\cite{1997GReGr..29.1527C},  which  is  close  to  our  present  study. As~ expected,  the~ authors  applied  their  methods  to  the  very  early  universe,  with~ the  Dirac  field  describing  massive  yet  microscopic  particles.  This  differs  from  our  present  study,  in~ which  we  assume  that  the  Dirac  field  describes  macroscopic  particles  within  a  static  universe.  }

For massive fermions, such as electrons and neutrons, the~mass will always appear in the phase factor of the solutions of wave equations. When gravity matters, this mass dependence remains. Consequently, it is generally believed that the gravity effect is not purely geometric at the quantum level. Hence, the immediate question is how can we reconcile, if~possible, the~contradiction between the quantum effect and the requirement of a geometric description of gravity? For microparticles, reconciling this contradiction is indeed impossible. However, in~this study, we propose that the scale-dependent quantum effect on macroscopic matter only becomes significant at cosmic scales. Locally, at~a macroscopic scale, the~large mass $m$ of each particle, or~equivalently, the~small value of the number density $\rho$ of the fluid, guarantees what is known as the classical limit, allowing for a geometric description of the gravitational effect on macroscopic matter. The~remaining question is what mechanism ensures that quantum effects are significant at cosmic scales while being negligible at macroscopic scales?

The answer to this question is not intricate, but~rather subtle. The~specifics will be provided later, but~for now, we will outline the main concepts behind the mechanism. Our analysis commences with an examination of the Dirac theory within various inertial frames~\citep{1973JMP....14..893H}. Generally, the~Lorentz--invariant Dirac spinor is defined in spacetime as
\begin{equation}\label{eq:general form of Dirac spinor}
\psi(x)=\rho^{1/2}e^{i\beta(x)/2}R(x).
\end{equation}
where $\rho(x)$ is a scalar representing the proper probability density and $R(x)$ is a rotor (Lorentz rotation) satisfying $R\tilde{R}=1$. In~Dirac theory, the~parameter $\beta(x)$ is intriguing. It is noteworthy that the states of a plane-wave particle have $\beta=0$, while those of an antiparticle have $\beta=\pi$. In~this paper, we aim to {generalize} the spinor $\psi(x)$ {to describe} free-falling macroscopic{, non-relativistic} particles {with identical mass $m$}. {Therefore,} it is logical to set $\beta=0$ throughout.  We {thus} adopt~\citep{Chen_2022}
\begin{equation}\label{eq:spinor}
  \psi(x)=\rho(x)^{1/2}R(x).
\end{equation}
{For} 
 our purpose, we can interpret $\rho=\psi\tilde{\psi}$ as the proper number density of a fluid, then we refer to $\rho_m=m\psi\tilde{\psi}$ as the corresponding proper mass density. As~illustrated in Appendix \ref{sec:ideal fluid}, the~rotor $R$ can be used to transform a fixed frame $\{\gamma_{\mu}\}$ into a new frame $\{e_{\mu}=R\gamma_{\mu}\tilde{R}\}$, and we identify $v=e_0$ as the proper velocity associated with the expected history $x(\tau)$ of a particle and thus the current velocity of the fluid. Naturally, the~corresponding  stress--energy tensor of the fluid can be written as
\begin{equation}\label{eq:T(a) for ideal fluid introduction}
T(a)=\rho_m a\cdot vv.
\end{equation}
It  is  important  to  point  out  that  this stress--energy  tensor,  which  originates  from  the  Dirac  spinor given by  Equation~(\ref{eq:spinor})  for  a  single  electron,  definitely  describes  a  classical  pressure-free  ideal  fluid  without  incorporating  quantum  spin.  Quantum  spin  is  only  included  if  the  spinor  satisfies  the  Dirac  equation,  as~ will  be  discussed  soon.  In~ this  sense,  we  can  say  that  the  spinor  given  in  Equation~(\ref{eq:spinor})  captures  all  aspects  of  macroscopic  particles,  when  considered with or without the  Dirac  equation.

The Dirac equation is
\begin{equation}\label{eq:Dirac equation in introduction}
\hslash\nabla\psi i\gamma_3=m\psi.
\end{equation}
where $\hslash$ is the Planck constant. This equation describes a single spin$-\frac{1}{2}$ free particle with a fixed mass $m$ and a probability density $\rho$. It is convenient to define a spin density trivector as (note that this is denoted with $S_3$ in Appendix \ref{sec:ideal fluid})
\begin{equation}\label{eq:spin trivector in introduction}
S=\frac{\hslash}{2}\psi i\gamma_3\tilde{\psi}=\frac{\hslash}{2}\rho R i\gamma_3\tilde{R}.
\end{equation}
We show in Appendix~\ref{sec:ideal fluid} that, {from the Dirac {Equation}
~(\ref{eq:Dirac equation in introduction})}, the~{general} stress--energy tensor for the spinor field is
\begin{equation}
T(a)= \rho_m a\cdot vv+[a\cdot\nabla(Sv)]\cdot v-(a\wedge\nabla)\cdot(Sv)v.
\end{equation}
When $S=0$ (this is the classical limit), this equation becomes the stress--energy tensor of a pressureless ideal fluid without spin, {as shown in  Equation~(\ref{eq:T(a) for ideal fluid introduction}). Clearly, if~we treat each free-falling macroscopic particle with a fixed mass $m$ as a single spin$-\frac{1}{2}$ free particle, we have the opportunity to include its quantum behavior under certain circumstances. This is the main aim of this study.} 

{We  are  now  in  a  position  to  discuss  how  the  spinor  approach  can  unify  quantum  mechanics  and  classical  mechanics.  We  believe  this  will  be  helpful  to  understand  the  physical  significance  of  our  proposal  in  this  study.  The~ general  form  of  the  Dirac  spinor,  as~ defined  in  Equation~(\ref{eq:general  form  of  Dirac  spinor}),  and~ the  corresponding  Dirac  equation  are  translations  of  the  traditional  matrix  version  of  Dirac  theory,  expressed  in  terms  of  STA~\cite{1973JMP....14..893H}.  They  are  equivalent  in  practical  applications.  The~ advantages  of  the  STA  version  of  Dirac  theory  include  providing  an  explicit  geometric  and  causal  interpretation  of  the  theory,  which  makes  it  very  convenient  when  incorporating  with  general  relativity  (GR).  One  of  the  unique  features  of  the  Dirac  equation  is  that  it  allows  for  solutions  with  both  positive  and  negative  energy  states.  The~ negative  energy  solutions  in  the  Dirac  equation  lead  to  the  concept  of  the  Dirac  sea,  a~ theoretical  model  that  was  used  to  explain  the  existence  of  antiparticles. Because~ of  the  presence  of  negative  energy  states,  when  dealing  with  Dirac  spinors,  one  encounters  different  types  of  physical  quantities  called  densities, such as scalar densities, which are Lorentz-invariant. Scalar  densities  can  interact  with  scalar  fields,  and~ the  Higgs  field  is  one  such  example, which  is  responsible  for  giving  mass  to  particles  through  their  interaction  with  it. When  we  generalize  the  Dirac  theory  to  describe  classical  macroscopic  particles, however,  we  restrict  ourselves  to  the  positive  energy  states.  This  is  adequate  because,  in~ the  case  of the non-relativistic  energy  of  macroscopic  particles,  there  are  no  particle--antiparticle  creation  processes  involved. That is, we treat each macroscopic particle as a single Dirac fermion. As~such, we recognize the proper number density $\rho=\psi\tilde{\psi}$ and the proper mass density $\rho_m=m\psi\tilde{\psi}$ (as measured by observers who are comoving with the current velocity $v=R\gamma_0\tilde{R}$) as that for the ensemble of a single macroscopic particle, described by the Dirac spinor give in Equation~(\ref{eq:spinor}) . As~mentioned, ~classical mechanics are recovered when $S=0$.}

When applying these results to gravitational systems, we simply need to replace $\nabla$ with the covariant derive symbol $D$, while all other quantities remain gauge-invariant.  As~such, the~classical limit (also known as the short wave approximation) refers to the fact that the contribution of the spin to the stress--energy is negligible. Clearly, the~magnitude of the spin $|S|=\frac{\hslash}{2}\rho$ is determined by the number density $\rho$. Therefore, an~alternative interpretation of the classical limit is that the value of the number density, $\rho$, is ignorably low, whereas a significant quantum effect arises when $\rho$ is sufficiently high. For~a fixed mass density $\rho_m(x)$, the~spin $|S|$ can be determined via $\rho(x)=\rho_m(x)/m$. This freedom strongly indicates that, in order to incorporate the scale-dependent quantum effect at scale, we can replace the mass $m$ in the Dirac equation with a mass function that depends on the spatial scale $\lambda$ of the system, denoted as $m(\lambda)$. The~mass function $m(\lambda)$ naturally satisfies the condition that, when $\lambda\rightarrow 0$, then $S\rightarrow 0$, corresponding to the macroscopic scale. Conversely, when $\lambda\rightarrow\infty$, then $S$ becomes a $\lambda$-dependent quantity related to the constant density of the universe. The~intermediate functional form of $m(\lambda)$, which plays a crucial role in the study of galaxies, can be determined through observations. Consequently, we can employ the Dirac equation in the presence of gravity without any modifications, as~there are no derivatives of $m(\lambda)$ involved in our calculations. The~properties of the mass function $m(\lambda)$ only need to be discussed in the final~results.

It is important to explicitly acknowledge that when applying the Dirac equation to gravitational systems at cosmic scales, we shall show in Section~\ref{sec:applications} that the anisotropy of the spin density adheres to the symmetric properties of mass density $\rho_m(x)$.  This fact only applies to macroscopic matter according to our assumption, where the gravitational effect is geometric throughout the spacetime manifold and, furthermore, the~proper mass density and the spin density have already been averaged and must satisfy the condition that the quantum effect vanishes locally. {In contrast}, when considering electrons, the~anisotropy of spin must be taken into account in all cases, as~the gravitational effect is not purely geometric. Thus, the quantum effect can never disappear locally, as~is commonly understood~\cite{1997GReGr..29.1527C,Brechet_2007,Brechet_2008,sakurai_napolitano_2017}.

The subsequent sections present an examination of radially symmetric and static gravitational systems based on our new postulation in Section~\ref{sec:radially symetric systems}, the~applications of our findings to cosmology and galaxies in Section~\ref{sec:applications}, and~ summarized conclusions and discussions in Section~\ref{sec:conclusions}. We employ natural units ($G=\hslash=c=1$) throughout, except where stated~otherwise.

\section{Radially Symmetric Gravitational~Systems}\label{sec:radially symetric systems}

We want to investigate the quantum nature of spacetime at cosmic scales, with~the main subjects of concern being galactic dynamics and cosmology. Traditionally, a~pressureless ideal fluid has been a good model for both galactic dynamics and matter-dominated era cosmology. In~both scenarios, matter experiences free-fall in a gravitational field and is characterized by the density distribution $\rho(x)$ and velocity field $v(x)$, where $x$ represents the position vector in spacetime. This model is usually known as collapsing dust. In~gravitational systems, it is widely recognized that a static state can only be sustained when there is a substantial gravitational potential well, as~observed in galaxies. However, in~the realm of cosmology, the~homogeneous and isotropic distribution of matter across the entire universe renders static solutions~non-existent.

Nevertheless, the~introduction of quantum effects at cosmic scales fundamentally alters the~situation.

The Dirac theory for radially symmetric gravitational systems can be investigated using gauge theory gravity (GTG).
The latter is constructed such that the gravitational effects are described by a pair of gauge fields, $\bar{h}(a)=\bar{h}(a,x)$ and $\omega(a)=\omega(a,x)$, defined over a flat Minkowski background spacetime~\citep{1998RSPTA.356..487L}, where $x$ is the STA position vector and is usually suppressed for short.
Luckily, the~majority of the necessary results for our present study have already been derived in prior work~\cite{1998JMP....39.3303D}. {These results are valid for microscopic particles; what we need to do is to generalize these to macroscopic matter. }

Let us consider a radially symmetric and static gravitational system composed of free-falling particles with the identical mass $m$. We make the assumption that the Dirac spinor
$\psi(r)=\rho(r)^{1/2}R(r)$ defined in (\ref{eq:spinor}) can fully capture all the essential physical aspects of the system if it satisfies the Dirac equation
\begin{equation}\label{eq:Dirac equation in gravity}
D\psi i\gamma_3=m\psi.
\end{equation}
We interpret $\rho=\psi\tilde{\psi}$ as the proper number density and  $\rho_m=m\psi\tilde{\psi}$ as the proper mass density. We adopt
$S=\frac{1}{2}\psi i\gamma_3\tilde{\psi}=\frac{1}{2}\rho R i\gamma_3\tilde{R}$ defined in (\ref{eq:spin trivector in introduction})
as the spin density trivector for the gravitational system. So there are four variables, namely $m$, $\rho$, $\rho_m$, and $S$, that can be used to describe the gravitational system. However, only two of them are independent, as~they must satisfy the conditions $\rho=\rho_m/m$ and $|S|=\frac{1}{2}\rho$. In~this paper, we eliminate $m$ and retain the other three variables. Among~them, the~relationship between $\rho_m$  and $\rho$ can be further determined through observations by assuming a specific form for the mass function $m(\lambda)$. Once this has been carried out, we are left with only one variable, which is the proper mass density $\rho_m$. Therefore, if~we are given the proper mass density $\rho_m$ of a gravitational system, we can predict the entire set of results, including the quantum~effects.

Now we define a set of spherical coordinates. From~the position vector of the \linebreak  flat~spacetime
\begin{equation}\label{eq:position vector}
x=t\gamma_0+r\sin\theta(\cos\phi\gamma_1+\sin\phi\gamma_2)+r\cos\theta \gamma_3
\end{equation},
we obtain the basis vectors, as follows:
\begin{equation}\label{eq:spherical basis}
\begin{split}
  e_t=\partial_t x=& \gamma_0, \\
  e_r=\partial_r x=&\sin\theta(\cos\phi \gamma_1+\sin\phi \gamma_2)+\cos\theta \gamma_3,  \\
  e_\theta=\partial_\theta x=&r\cos\theta(\cos\phi \gamma_1+\sin\phi \gamma_2)-r\sin\theta \gamma_3, \\
  e_\phi=\partial_\phi x=&r\sin\theta(-\sin\phi \gamma_1+\cos\phi \gamma_2).
\end{split}
\end{equation}
Since $e_\theta$ and $e_\phi$ are not unit, we define
\begin{equation}\label{eq:unit basis}
\hat{\theta}\equiv e_\theta/r, \; \; \; \hat{\phi}\equiv e_\phi/(r\sin\theta).
\end{equation}
With these unit vectors, we further define the unit bivectors (relative basis vectors for $e_t=\gamma_0$)
\begin{equation}\label{eq:pauli basis}
\begin{split}
   \sigma_r\equiv &e_re_t, \\
     \sigma_\theta\equiv &\hat{\theta}e_t, \\
     \sigma_\phi\equiv &\hat{\phi}e_t.
\end{split}
\end{equation}
These bivectors satisfy
\begin{equation}\label{eq:pseudo-scalar}
\sigma_r\sigma_\theta\sigma_\phi=e_te_r\hat{\theta}\hat{\phi}=i.
\end{equation}

As in our previous paper ~\citep{Chen_2022}, we initially attempted to analyze static systems. The~set of $\bar{h}$ field that satisfies the spherically symmetric and static matter distribution is assumed to take the form~\citep{1998RSPTA.356..487L,doran2003geometric}
\begin{equation}\label{eq:h function}
\begin{aligned}
   &\bar{h}(e^t)=f_1e^t,   &&\bar{h}(e^r)=g_1e^r+g_2e^t,  \\
   &\bar{h}(e^{\theta})=e^{\theta},  &&\bar{h}(e^{\phi})=e^{\phi},
\end{aligned}
\end{equation}
where $f_1$, $g_1$, and $g_2$ are all functions of $r$ only. We could have tried the form $\bar{h}(e^t)=f_1e^t+f_2e^r$, but~it is more reasonable to set $f_2$ to zero, which is referred to as the `Newtonian gauge'~\citep{1998RSPTA.356..487L}.

The subsequent steps of this study can be summarized as follows:
\begin{center}
  ~\\
$\bar{h}(e^{\mu})\xrightarrow{D\wedge\bar{h}(a)=\kappa \bar{h}(a)\cdot S} \omega(a)$

~\\
\noindent $\xrightarrow{\mathcal{R}(a\wedge b)=L_a\omega(b)-L_b\omega(a)+\omega(a)\times\omega(b)}$
$\begin{Bmatrix}
 \mathcal{R}(a)=\partial_b\cdot\mathcal{R}(b\wedge a)& \\
 \mathcal{R}=\partial_a\cdot\mathcal{R}(a)&
 \end{Bmatrix}
$

~\\

\noindent $\xrightarrow{\mathcal{G}(a)=\kappa\mathcal{T}(a)}g_{\mu\nu}=\underline{h}^{-1}(\mu)\cdot\underline{h}^{-1}(\nu),
$
~\\
\end{center}

\noindent where the metric $g_{\mu\nu}$ is introduced to compare our results with those predicted by general relativity (GR) and to provide conventional approaches for subsequent applications. Notably, $\omega(a)$ and $\mathcal{R}(a\wedge b)$ can be decomposed into torsion-free components and torsion components~\cite{1998JMP....39.3303D}. This decomposition allows for the recovery of classical predictions of GR when the impact of torsion is insignificant. Conversely, given the extensive research on these classical predictions available in the existing literature, we can easily incorporate torsion terms into classical results when we deem them to be significant. This allows us to leverage decompositions and enhance our understanding of the phenomena under~consideration.

By solving Equation~(\ref{eq:torsion}), we can obtain a solution for $\omega(a)$~\citep{1998JMP....39.3303D,1998RSPTA.356..487L}, the~result is
\begin{equation}\label{eq:omega text}
 \omega(a)=\omega'(a)+\frac{1}{2}\kappa a\cdot S,
\end{equation}
where
\begin{equation}\label{eq:omega prime as a function of a}
\begin{aligned}
\omega'(a)=&(a\cdot e_t G-a\cdot e_r F)e_re_t-\left(\frac{g_2}{r}\right)a\cdot\hat{\theta}\hat{\theta}e_t \\
 & -\left(\frac{g_1-1}{r}\right)a\cdot\hat{\theta}e_r\hat{\theta}-\left(\frac{g_2}{r}\right)a\cdot\hat{\phi}\hat{\phi}e_t \\
 & -\left(\frac{g_1-1}{r}\right)a\cdot\hat{\phi}e_r\hat{\phi}
\end{aligned}
\end{equation}
denotes the torsion-free component of $\omega(a)$, and~the new functions $G$ and $F$ are also all functions of $r$ only. From~the $\omega$ field, for~any bivector $B$, its strength tensor can be obtained directly from Equation~(\ref{eq:curvature})~\citep{1998RSPTA.356..487L,1998JMP....39.3303D}, the~result is
\begin{equation}\label{eq:R(a b)}
\begin{split}
\mathcal{R}(B)=& \mathcal{R}'(B)+\frac{1}{4}\kappa^2(B\cdot S)\cdot S  \\
     & -\frac{1}{2}\kappa(B\cdot D)\cdot S,
\end{split}
\end{equation}
where
\begin{equation}\label{eq:R'B}
\begin{split}
   \mathcal{R}'(B)= & \alpha_1\sigma_r B\cdot\sigma_r+(\alpha_2\sigma_\theta+\alpha_3i\sigma_\phi)B\cdot\sigma_\theta \\
     & +(\alpha_2\sigma_\phi-\alpha_3i\sigma_\theta)B\cdot\sigma_\phi+\alpha_6\sigma_r(B\wedge\sigma_r) \\
     & +(\alpha_4\sigma_\theta-\alpha_5i\sigma_\phi)(B\wedge\sigma_\theta) \\
     & +(\alpha_4\sigma_\phi+\alpha_5i\sigma_\theta)(B\wedge\sigma_\phi)
\end{split}
\end{equation}
denotes the torsion-free component of $R(B)$, with~the understanding that $\sigma_r\wedge\sigma_\theta=\sigma_\phi\wedge\sigma_\theta=0$, and~$\alpha_1, \ldots, \alpha_6$ are given by
\begin{equation}\label{eq:alphas}
\begin{split}
  \alpha_1=&L_rG-L_tF+G^2-F^2, \\
  \alpha_2=&-L_t\left(\frac{g_2}{r}\right)+\left(\frac{g_1}{r}\right)G-\left(\frac{g_2}{r}\right)^2,  \\
  \alpha_3=&L_t\left(\frac{g_1}{r}\right)+\frac{g_1g_2}{r^2}-\left(\frac{g_2}{r}\right)G, \\ \alpha_4=&L_r\left(\frac{g_1}{r}\right)+\left(\frac{g_1}{r}\right)^2-\left(\frac{g_2}{r}\right)F, \\
  \alpha_5=&L_r\left(\frac{g_2}{r}\right)+\frac{g_1g_2}{r^2}-\left(\frac{g_1}{r}\right)F, \\
  \alpha_6=&(-g_2^2+g_1^2-1)/r^2.
\end{split}
\end{equation}
From Equations~(\ref{eq:R(a b)}) and (\ref{eq:R'B}), the~Ricci tensor $\mathcal{R}(a)$ and Ricci scalar $\mathcal{R}$ are given by
\begin{equation}\label{eq:R(a)and R}
\begin{split}
   \mathcal{R}(a)=& R'(a)+\frac{1}{2}\kappa^2(a\cdot S)\cdot S  \\
        &-\frac{1}{2}\kappa a\cdot(D\cdot S), \\
  \mathcal{R}=& \mathcal{R}'+\frac{3}{2}\kappa^2S^2,
\end{split}
\end{equation}
where
\begin{equation}\label{eq:R'a}
\begin{split}
   \mathcal{R}'(a)= & [(\alpha_1+2\alpha_2)a\cdot e_t+2\alpha_5a\cdot e_r]e_t \\
     & +[2\alpha_3a\cdot e_t-(\alpha_1+2\alpha_4)a\cdot e_r]e_r \\
     & -(\alpha_2+\alpha_4+\alpha_6)a\cdot\hat{\theta}\hat{\theta} \\
     & -(\alpha_2+\alpha_4+\alpha_6)a\cdot\hat{\phi}\hat{\phi},
\end{split}
\end{equation}
and
\begin{equation}\label{eq:R'}
\mathcal{R}'=2\alpha_1+4\alpha_2+4\alpha_4+2\alpha_6,
\end{equation}
denote the torsion-free components of $\mathcal{R}(a)$ and $\mathcal{R}$, respectively.

In order to solve Einstein Equation~(\ref{eq:Einstein equation}), we need to know $T(a)$ given by Equation~(\ref{eq:energy-momentum tensor}). Noting that $\psi=\psi(r)$, we find that
\begin{equation}\label{eq:stress-energy tensor in gravity}
\begin{split}
   \mathcal{T}(a)=&\langle a\cdot D\psi i\gamma_3\tilde{\psi}\rangle_1 \\
     =&\langle a\cdot\bar{h}(e^{\mu})\partial_{\mu}\psi i\gamma_3\tilde{\psi}+\frac{1}{2}\omega(a)\psi i\gamma_3\tilde{\psi}\rangle_1 \\
     =&a\cdot(g_1e^r+g_2e^t)\langle\partial_r\psi i\gamma_3\tilde{\psi}\rangle_1+\omega(a)\cdot S.
\end{split}
\end{equation}
The terms $\langle\partial_r\psi i\gamma_3\tilde{\psi}\rangle_1$ should be derived from the Dirac Equation~(\ref{eq:Dirac equation in gravity}).  Nevertheless, as~elaborated in our previous paper~\cite{Chen_2022},  the~use of this stress--energy tensor prototype can be misleading when applied to gravitational systems consisting of macroscopic bodies. This is because, as~$S$ approaches zero, the~tensor $T(a)$ defined in (\ref{eq:stress-energy tensor in gravity}) also approaches zero. This contradicts our expectation that as $S$ tends to zero, $T(a)$ should represent a tensor describing a {classical} pressureless ideal~fluid.

An appropriate representation of the stress--energy tensor for our specific needs involves decomposing it into two components: one that characterizes the classical pressure-free ideal fluid and another that accounts for quantum effects. By~substituting $\nabla$ with $D$, we can rephrase Equation~(\ref{eq:coordinate-free of adjoint of T}) and express $\mathcal{T}(a)$ as,
\begin{equation}\label{eq:decomposition of T(a)}
\mathcal{T}(a)=\rho_m a\cdot e_te_t-(a\wedge D)\cdot(S\cdot e_t)e_t+[a\cdot D(S\cdot e_t)]\cdot e_t,
\end{equation}
where we have replaced $v$ with $e_t$ due to the retained gauge freedom to perform arbitrary radial boosts in restricting the $\bar{h}$ function~\citep{1998RSPTA.356..487L,doran2003geometric}. {It is important to note that the choice $v=e_t=\gamma_0$ simply fixes the rotation gauge in such a way that the stress--energy tensor takes on the simplest form; there is no other additional physical content. Furthermore, setting $v=e_t$ does not imply that the fluid is at rest or that the observers are comoving with the fluid. In~such a setting, all rotation-gauge freedom has been completely removed, as~it should be, before~one can derive a complete set of physical equations. Note also that $\mathcal{R}(B)$ deals directly with physically measurable quantities, whereas the algebraic structure of the $h$-function is of little direct physical significance. Therefore, the~rotation gauge has been fixed by imposing a suitable form for $\mathcal{R}(B)$, rather than restricting the form of $\bar{h}(a)$, as~discussed in Ref.~\citep{1998RSPTA.356..487L,doran2003geometric}. Consequently, the~proper mass density $\rho_m(r)$, potential energy density $\rho_Q(r)$, and the corresponding mass (energy) $M(r)$ presented later acquire a similar physical meaning as in GR. }

The Einstein equation is
\begin{equation}\label{eq:Einstein equation specific}
\mathcal{G}(a)=R(a)-\frac{1}{2}a\mathcal{R}=\kappa\mathcal{T}(a).
\end{equation}
A direct and efficient approach is to solve the Einstein equation separately for two scenarios: when $a=e_t$ and $a=e_r$. To~be specific, in~general, our results need to be derived from the following equations:
\begin{equation}\label{eq:four Einstein equations}
\begin{split}
   e_t\cdot \mathcal{G}(e_t)=&\kappa e_t\cdot \mathcal{T}(e_t),  \\
    e_r\cdot \mathcal{G}(e_t)=&\kappa e_r\cdot \mathcal{T}(e_t), \\
    e_t\cdot \mathcal{G}(e_r)=&\kappa e_t\cdot \mathcal{T}(e_r), \\
    e_r\cdot \mathcal{G}(e_r)=&\kappa e_r\cdot \mathcal{T}(e_r).
\end{split}
\end{equation}
However, we demonstrate that, for~our intended purpose, solving the first two equations listed above is adequate. From~(\ref{eq:decomposition of T(a)}), we obtain
\begin{equation}\label{eq:e_t dot T(e_t) annother form}
e_t\cdot\mathcal{T}(e_t)=\rho_m-(e_t\wedge D)\cdot(S\cdot e_t).
\end{equation}
Since $(e_t\wedge D)\cdot(S\cdot e_t)=(e_t\wedge\partial_a)\cdot[a\cdot D(S\cdot e_t)]$, we need to calculate $a\cdot D(S\cdot e_t)$, as follows:
\begin{equation}\label{eq:a.D(S.et)}
\begin{split}
   a\cdot D(S\cdot e_t) =& (a\cdot DS)\cdot e_t+S\cdot(a\cdot De_t) \\
     =& [a\cdot\bar{h}(e^r)\partial_r S+\omega(a)\times S]\cdot e_t+S\cdot[\omega(a)\cdot e_t].
\end{split}
\end{equation}
We thus have
\begin{equation}\label{eq:et wedge D dot S dot et}
\begin{split}
   (e_t\wedge\partial_a)\cdot[a\cdot D(S\cdot e_t)]= & (e_t\wedge\partial_a)\cdot[S\cdot(\omega(a)\cdot e_t)] \\
    = & e_t\cdot\{[\partial_a\wedge(\omega(a)\cdot e_t)]\cdot S\} \\
    = & e_t\cdot\{[-G\sigma_r+\kappa(S\cdot e_t)]\cdot S\} \\
    = & \kappa (e_t\cdot S)^2=\kappa S^2.
\end{split}
\end{equation}
Therefore,
\begin{equation}\label{eq:et dot T et final}
e_t\cdot\mathcal{T}(e_t)=\rho_m-\kappa S^2.
\end{equation}
Similarly, we have
\begin{equation}\label{eq:er dot T et}
\begin{split}
   e_r\cdot\mathcal{T}(e_t)= & [e_t\cdot D(S\cdot e_t)]\cdot(e_t\wedge e_r) \\
     = & G(S\cdot e_r)\cdot(e_t\wedge e_r) \\
     = & 0.
\end{split}
\end{equation}

On the other hand, for~$\mathcal{G}(e_t)$, we find from (\ref{eq:R(a)and R})--(\ref{eq:R'}) that
\begin{equation}\label{eq:G(e_t)}
\begin{split}
   \mathcal{G}(e_t) =& R(e_t)-\frac{1}{2}e_t R \\
     =& 2\alpha_3e_r-(2\alpha_4+\alpha_6+\frac{3}{4}\kappa^2S^2)e_t \\
     &+\frac{1}{2}\kappa^2(e_t\cdot S)\cdot S-\frac{1}{2}\kappa e_t\cdot(D\cdot S).
\end{split}
\end{equation}
Substituting (\ref{eq:et dot T et final})--(\ref{eq:G(e_t)}) into $e_t\cdot\mathcal{G}(e_t)=\kappa e_t\cdot\mathcal{T}(e_t)$ and $e_r\cdot \mathcal{G}(e_t)=\kappa e_r\cdot \mathcal{T}(e_t)$, we obtain
\begin{equation}\label{eq: the results from the first two Einstein eqs}
\begin{array}{ccl}
2\alpha_4+\alpha_6&=&-\kappa(\rho_m-\frac{3}{4}\kappa S^2) \\
\alpha_3&=&0.
\end{array}
\end{equation}
Interestingly, from~(\ref{eq:R'a}) we find
\begin{equation}\label{eq:partial_a wedge R'a}
\partial_a\wedge \mathcal{R}'(a)=2\alpha_5\sigma_r-2\alpha_3\sigma_r=0,
\end{equation}
which gives
\begin{equation}\label{eq:alpha_5}
\alpha_5=\alpha_3=0.
\end{equation}

By substituting the expressions of $\alpha$s given by (\ref{eq:alphas}) into (\ref{eq: the results from the first two Einstein eqs}) and (\ref{eq:alpha_5}), our solutions to the Einstein equations can be summarized as follows:
\begin{equation}\label{eq:alphas replaced for solutions to Einstein equation}
\begin{split}
  2\left[ g_1\partial_r(\frac{g_1}{r})+(\frac{g_1}{r})^2-F(\frac{g_2}{r})\right]-\frac{g_2^2-g_1^2+1}{r^2} = &-\kappa(\rho_m-\frac{3}{4}\kappa S^2), \\
  g_2\partial_r(\frac{g_1}{r})+\frac{g_1g_2}{r^2}-G\frac{g_2}{r} = & 0, \\
  g_1\partial_r(\frac{g_2}{r})+\frac{g_1g_2}{r^2}-F\frac{g_1}{r} = & 0.
\end{split}
\end{equation}

The first and third equations in (\ref{eq:alphas replaced for solutions to Einstein equation}) can be combined to give
\begin{equation}\label{eq:the pre-mass equation}
\partial_r\left[r\left((g_2)^2-(g_1)^2+1\right)\right]=\kappa\left(\rho_m-\frac{3}{4}\kappa S^2\right)r^2.
\end{equation}
Now, if we define
\begin{equation}\label{eq:defining mass}
M=\frac{r}{2}\left((g_2)^2-(g_1)^2+1\right),
\end{equation}
we find (remembering $\kappa=8\pi$):
\begin{equation}\label{eq:mass expression}
M(r)=4\pi\int_{0}^{r}\left(\rho_m(r')-\frac{3}{4}\kappa S^2(r')\right)r'^2 d r'.
\end{equation}
The expression of $M$ strongly implies that it can be identified as the mass (total energy) of a gravitational system within $r$ at any given time $t$. Remarkably, the~mass $M$ contains the quantum effects, $-\frac{3}{4}\kappa S^2(r)$, as~intended. Naturally, when $S=0$, the~conventional expression for $M$ in general relativity is~regained.

As mentioned earlier, there are three functions in the $\bar{h}$ field, namely $f_1$, $g_1$, and $g_2$, that need to be determined by solving Einstein equations. So far, we have obtained expressions for $g_1$ and $g_2$, and they are linked to the observables $\rho_m$ and $S$. Thus, the~determination of $f_1$ remains pending. To~achieve this, we substitute $a=e^t$ into the torsion Equation~(\ref{eq:torsion}), yielding the subsequent equations
\begin{equation}\label{eq:torsion equation for et}
\begin{split}
   D\wedge \bar{h}(e^t) =& \kappa\bar{h}(e^t)\cdot S, \\
     \partial_a\wedge\left[a\cdot\bar{h}(e^{\mu})\partial_{\mu}f_1e^t+f_1\omega(a)\cdot e^t\right] =&\kappa f_1e_t\cdot S, \\
   g_1\partial_rf_1=& -Gf_1,  \\
     f_1=& e^{-\int\frac{G}{g_1}dr}.
\end{split}
\end{equation}
On the other hand, from~the last two equations of (\ref{eq:alphas replaced for solutions to Einstein equation}), it is easy to derive that
\begin{equation}\label{eq:F and G}
G=\partial_r g_1, \ \ F=\partial_rg_2.
\end{equation}
So, we immediately obtain that
\begin{equation}\label{eq:f_1}
f_1=1/g_1.
\end{equation}

Now, we turn to the metric tensor $g_{\mu\nu}=g_{\mu}\cdot g_{\nu}=\underline{h}^{-1}(e_{\mu})\cdot\underline{h}^{-1}(e_{\nu})$. From~(\ref{eq:h function}) and~(\ref{eq:f_1}), we obtain
\begin{equation}\label{eq:underline{h}^{-1}}
\begin{aligned}
&\underline{h}^{-1}(e_t)=g_1e_t-g_2e_r, &&\underline{h}^{-1}(e_r)=\frac{1}{g_1}e_r, \\
&\underline{h}^{-1}(e_{\theta})=e_{\theta}, &&\underline{h}^{-1}(e_{\phi})=e_{\phi}.
\end{aligned}
\end{equation}
So, the metric tensor depends only on $g_1$ and $g_2$. However, the $(g_2)^2$ included in (\ref{eq:defining mass}) served as the kinematic energy in gravitational systems~\cite{1998RSPTA.356..487L}. To~see this, we consider a radially free-falling particle with $v=e_t$. We have
\begin{equation}\label{eq:free fall particle}
\dot{x}=\frac{dx}{d\tau}=\dot{t}e_t+\dot{r}e_r=\underline{h}(v)=\underline{h}(e_t)=\frac{1}{g_1}e_t+g_2e_r.
\end{equation}
Clearly, in~this case, $g_2=\dot{r}$ represents the radial velocity of the particle. In~general, we can understand the physical significance of $g_2$ by rewriting (\ref{eq:defining mass}) as~\cite{1998RSPTA.356..487L}
\begin{equation}\label{eq:Bernoulli equation}
\frac{1}{2}(g_2)^2-\frac{M}{r}=\frac{1}{2}\left((g_1)^2-1\right),
\end{equation}
which is a Bernoulli equation for zero pressure and total non-relativistic energy $\frac{1}{2}((g_1)^2-1)$.

But for non-relativistic matter, the~contribution of kinematic energy to gravity can be safely disregarded. So, from (\ref{eq:defining mass}) we have
\begin{equation}\label{eq:g_1 and M}
g_1=\left(1-\frac{2M(r)}{r}\right)^{1/2}.
\end{equation}
This paper specifically concentrates on non-relativistic matter within static gravitational systems, enabling us to express the metric as follows:
\begin{equation}\label{eq:interval in spacetime}
d\tau^2=\left(1-\frac{2M(r)}{r}\right)dt^2-\left(1-\frac{2M(r)}{r}\right)^{-1}dr^2-r^2(d\theta^2+\sin^2\theta d\phi^2).
\end{equation}
Remarkably, we have obtained a metric that bears a striking resemblance to the familiar form in general relativity. However, it is important to note that this metric was derived from Dirac theory, and~the mass $M$ incorporates the quantum effects arising from macroscopic~matter.

In this paper, the~term $-\frac{3}{4}\kappa S^2(r)$ in the mass--energy expression (\ref{eq:mass expression}) is referred to as the ``quantum potential energy''. This concept was first introduced by Bohm~\cite{Bohm1952} in 1952 and later by DeWitt~\cite{DeWitt1952} in the same year, with~both authors commonly referring to it as the ``quantum potential.''~\cite{sym16010067}. Similarly to how traditional mass--energy curves spacetime, quantum potential energy distorts spacetime. This phenomenon is known as ``spin-torsion'' theory~\cite{1976RevModPhys.48.393,1998RSPTA.356..487L}. By~explicitly decomposing the stress--energy tensor $\mathcal{T}(a)$ into a spin-free part and a spin part, as~illustrated in (\ref{eq:decomposition of T(a)}), the~gravitational strength, represented by the Riemann tensor $\mathcal{R}(B)$, can be explicitly decomposed into a torsion-free component and a torsion component, as shown in (\ref{eq:R(a b)}). This results in a two-component metric as outlined in (\ref{eq:interval in spacetime}). Naturally, when $S=0$, the~metric reduces to one that describes a spherically-static gravitational system, as suggested by~GR.

It is crucial to grasp the physical significance of the quantum potential energy (QPE) in this metric. To date, in the literature, QPE has been attributed uniquely to microparticles.
It has been confirmed that QPE would reduce to a time dilation in spacetime. For~instance, when negative muons are captured in atomic s-states, their lifetimes are increased by a time dilation factor corresponding to the Bohr velocity. So, we expect that, if~this QPE-induced time dilation is extended to macroscopic matter at cosmic scales, it may be disguised as some extra gravitational potential well. We will show that this extra time dilation can successfully explain the cosmic redshift and galactic rotation curve problem. For~microparticles, the~troubles encountered all arise from the fact that when the QPE is significant, its gravitational effect is not geometric (i.e., mass-dependent). This results in the concept of the geodesic for test particles becoming ambiguous. Therefore, as~mentioned earlier, if~we want to apply this metric to gravitational systems composed of macroscopic particles, so that QPE is significant only at cosmic scales and can be neglected locally, we must establish a mechanism that connects the quantum effects of macroscopic matter to the spatial scales of gravitational systems.
~The~mechanism, however, can be clearly demonstrated in the applications of our theory to cosmology and galaxies, as~illustrated in the following~section.

\section{Applications: Cosmology and Galaxies}\label{sec:applications}

To investigate the mechanism that connects the quantum effects of macroscopic matter to the spatial scales of gravitational systems, we begin by considering a radially symmetric and static gravitational system consisting of $N$ free-falling particles, each with an identical mass of $m$. We assume that the direct collisions and the close encounters due to gravity between the particles can be neglected. This allows us to approximate the gravitational effect of the particles by a smooth distribution of matter.
Namely, we assume a smooth mass density function $\rho_m(r)$ and a smooth number density function $\rho(r)$ defined by
\begin{equation}\label{eq: rho_m and rho}
\rho_m(r)=m\rho(r)=\sum_{i=1}^{N}m\delta(r-r_i).
\end{equation}
It is evident that the accuracy of the approximation improves as the mass $m$ decreases and the particle number $N$ increases. However, for~a fixed mass density $\rho_m(r)$, reducing the mass $m$ leads to an increase in the number density $\rho(r)$. On~the other hand, the~scale-dependence of a particle's mass in a gravitational system can be easily demonstrated.  A~particle with a mass of $m$ contributes to the average mass density of a system with a size of $\lambda$ through the following relation:
\begin{equation}\label{eq:scale dependence of a particle's mass}
\rho(m,\lambda)\sim \frac{m}{\lambda^3}.
\end{equation}
Clearly, from~the perspective of average mass density, the~``effective mass'' of a particle decreases as the size of the system increases, following a trend of $\sim$$\lambda^{-3}$. While this may seem trivial, it becomes crucial in our understanding of quantum effects at cosmic scales. For~example, in~our Milky Way, $\lambda$$\sim$10 kpc. The~Sun ($m=M_{\odot}$) contributes to the average density of the Milky Way as an electron ($m=9.1\times10^{-28}{}{g}$) {contributes} 
 to a system of size $\lambda$$\sim$230 cm, a~macroscopic size. So, as an assumption, we extend the quantum nature of electrons to macroscopic matter particles distributed in sufficiently large ~systems.

In a gravitational system, the~Dirac equation for a free-falling particle describes a particle of fixed {proper} mass $m$,  a~constant spin $\hslash/2$, and a {proper} probability density $\rho(r)$. We consider the probability density as the number density of the system. In~the case of spin-$\frac{1}{2}$ microparticles, the~mass $m$ can vary from neutrinos ($m\approx0$) to neutrons, while the spin remains constant at $\hslash/2$. Therefore, the~spin of a particle is independent of its mass. This fact can be naturally extended to macroscopic particles. Typically, when dealing with large masses (macroscopic matter), the~negligible quantum effect can be understood as the short wavelength approximation, commonly referred to as the classical limit. In~the context of radially symmetric gravitational systems, we can naturally interpret the negligible quantum effect as a result of the negligibly small number density of particles (since each particle has exactly a fixed value $\hslash/2$ of spin). Given that reducing the mass $m$ results in an increase in the number density $\rho(r)$, we can introduce a decreasing function $m(\lambda)$ ($\lambda$ is the size of the gravitating system), which in turn leads to an increasing function $\rho(r)$. Consequently, the~strength of the quantum effect would increase as the distance (or scale) to an observer increases, aligning with our~postulation.

To be more precise, we define the density of QPE from (\ref{eq:mass expression}) as (note that $S^2=-\frac{1}{4}\rho^2$, and~$\rho(r)=\rho_m(r)/m(\lambda)$)
\begin{equation}\label{eq:rho_Q}
\begin{split}
\rho_Q(r,\lambda)=&-\frac{3}{4}\kappa S^2(r) \\
         =&\frac{3}{16}\kappa\rho(r)^2 \\
         =&\frac{3}{16}\kappa\left(\frac{\rho_m(r)}{m(\lambda)}\right)^2 \\
         =&\frac{\rho_m^2(r)}{\rho_c}\alpha(\lambda),
\end{split}
\end{equation}
where $\alpha(\lambda)$ is a dimensionless parameter that depends on the scale $\lambda$. We require that the strength of the QPE vanishes when $\lambda\rightarrow 0$ and increases with increasing $\lambda$. Note that, in order to  separate the size $\lambda$ dependence of $\rho_Q$ completely from coordinate $r$, we express $\rho_Q(r,\lambda)$ as proportional to $\rho_m^2(r)$, and~the constant mass density $\rho_c$ of the entire universe is introduced into the definition of $\rho_Q$ only for dimensional consistency. This choice has nothing to do with our contemplation of the notion that the entire universe may exert a cosmological influence on local galaxies, as~elaborated upon in subsequent sections. Needless to say, the~functional form of $\alpha(\lambda)$ must be universal, meaning that it is independent of any particular gravitational system, although~it can be different for the entire universe and galaxies. {The specific properties of $\alpha(\lambda)$ should  be  determined  theoretically  from  first  principles.  However,  as~ a  first  attempt  to  compare  our  predicted  results  with  observations,  when  theoretical  formulas  for $\alpha(\lambda)$  are  lacking,  we  can  derive  a  phenomenological  formula  by  fitting  the  data  to  observations.}

From (\ref{eq:mass expression}), we observe that the expression for $M$ can be rewritten as
\begin{equation}\label{eq:M and rhoQ}
\begin{split}
   M(r)= & 4\pi\int_{0}^{r}(\rho_m(r')+\rho_Q(r',\lambda))r'^2dr' \\
       = & 4\pi\int_{0}^{r}\rho_m(r')r'^2dr'+\frac{4\pi\alpha(r)}{\rho_c}\int_{0}^{r}\rho_m^2(r')r'^2dr' \\
       = & M_m(r)+M_Q(r),
\end{split}
\end{equation}
where we have set $\lambda=r$ in the integrand to reflect the fact that, in~$\rho_Q(r',\lambda)$, the~scale $\lambda$ is  $r$, the~size of the ``system'' with mass $M(r)$. And~\begin{eqnarray}
  M_m(r) &=& 4\pi\int_{0}^{r}\rho_m(r')r'^2dr' \\
  M_Q(r) &=& \frac{4\pi\alpha(r)}{\rho_c}\int_{0}^{r}\rho_m^2(r')r'^2dr'
\end{eqnarray}
represent the conventional mass (energy) in general relativity and the QPE within $r$, respectively. From~(\ref{eq:rho_Q}) and (\ref{eq:M and rhoQ}), it can be clearly seen that the distribution of QPE exhibits no extra symmetries, i.e.,~it aligns with the mass distribution. The~reason for this is that the scale $\lambda$ characterizes only the amount of QPE for macroscopic matter at cosmic scales. Therefore, as~the scale increases, the~amount of QPE also increases, causing the quantum effects to vanish locally. This is in contrast to electrons, whose QPE can never vanish at any scale,~thus resulting in possible anisotropy if not averaged~\cite{Brechet_2007,Brechet_2008}.

As previously mentioned, the~functional form of $\alpha(\lambda)$ should be determined based on observations. Nonetheless, we have found that the following general form is a suitable~choice:
\begin{equation}\label{eq:functional form alpha}
\alpha(x)=\frac{A(e^x-1)}{x^2}+B(e^x-1)+C, \ \  \ \text{with} \ \ \  x=H_0\lambda,
\end{equation}
where $H_0$ represents the Hubble constant, which will be defined subsequently. We leave the coefficients $A$, $B$, and $C$, possibly scale dependent, as~free parameters to fit the observational data of the gravitational systems under~consideration.

\subsection{Cosmology}
Now we are ready to study cosmology. Assuming a static universe with a homogeneous and isotropic matter distribution at large scales, characterized by a constant mass density $\rho_c$, the~mass $M$ in (\ref{eq:M and rhoQ}) then becomes
\begin{equation}\label{eq:M for cosmology}
\begin{split}
M(r)=&4\pi\int_{0}^{r}\rho_cr'^2dr'+4\pi\alpha(r)\int_{0}^{r}\rho_c r'^2dr' \\
    =&\frac{1}{2}H_0^2r^3(1+\alpha(r)),
\end{split}
\end{equation}
where
\begin{equation}\label{eq:H_0}
H_0^2=\frac{8\pi\rho_c}{3}
\end{equation}
is the Hubble constant. In~cosmology, the~entire universe can be regarded as an gravitational system with an arbitrarily large scale. To~compare observations, we suggest setting the parameters in (\ref{eq:functional form alpha}) as $A=2$, $B=0$, and $C=-1$. Namely, for~cosmology, we assume
\begin{equation}\label{eq:cosmic alpha}
\alpha_c(\lambda)=\frac{2(e^{H_0\lambda}-1)}{H_0^2\lambda^2}-1,
\end{equation}
which yields desirable results:
\begin{equation}\label{eq:M for cosmology with special alpha}
M(r)=r(e^{H_0r}-1)=H_0r^2+\frac{1}{2}H_0^2r^3+\cdots.
\end{equation}
Before formulating the metric for the spacetime of the universe, it is important to highlight that the term $\frac{1}{2}H_0^2r^3$ resulting from $\rho_c$ in $M(r)$ given in (\ref{eq:M for cosmology}) should not be disregarded. One could argue that the universe is a distinct gravitational system, exhibiting a homogeneous and isotropic distribution of matter, and~infinite in size. Consequently, the~observation of gravitational redshift may not be possible under these circumstances. However, this conclusion is derived from Newtonian theory or general relativity, and~does not hold true when quantum effects come into play. In~fact, the~term $\frac{1}{2}H_0^2r^3$ can be included as the QPE if we choose $C=0$ in (\ref{eq:functional form alpha}). In~any case, we should keep in mind that any choice of $\alpha_c(\lambda)$ must be subjected to testing against observations. Moreover, due to the presence of quantum effects at cosmic scales, we have to abandon the idea of the absolute spacetime predicted by general relativity. This is quite similar to the case of special relativity, when Einstein had to relinquish the concept of Newtonian absolute space and time. As~a result, the~conventional notion of the geometry of the entire universe becomes meaningless. For~instance, it loses significance to classify the universe as open, flat, or~closed according to its mass density $\rho_c$. Thus, for cosmology, we have from (\ref{eq:g_1 and M})
\begin{equation}\label{eq:g1 for cosmology}
g_1=\left(1-\frac{2M(r)}{r}\right)^{1/2}=(3-2e^{H_0r})^{1/2}.
\end{equation}

Hence, the metric (\ref{eq:interval in spacetime}) for the universe becomes
\begin{equation}\label{eq:metric for cosmology}
\begin{split}
d\tau^2=&(3-2e^{H_0r})dt^2-(3-2e^{H_0r})^{-1}dr^2 \\
        &-r^2(d\theta^2+\sin^2\theta d\phi^2).
\end{split}
\end{equation}
From this metric, it can be readily deduced that a horizon exists in the static universe, with~the distance to any observer in the universe given by
\begin{equation}\label{eq:horizon}
r_h=\frac{1}{H_0}\ln(3/2)\approx\frac{0.4}{H_0},
\end{equation}
which is less than half a Hubble radius $1/H_0$.

As a crucial outcome of our theory, we now derive the Hubble redshift for a static universe model. Suppose a light signal emitted from a source at $r$ and received by an observer at $r=0$, the~redshift arising from the time dilation can be readily derived as
\begin{equation}\label{eq:redshift}
1+z=\frac{g_1(0)}{g_1(r)}=\frac{1}{\sqrt{3-2e^{H_0r}}}.
\end{equation}
It is evident that the redshift approaches infinity as the light source approaches the horizon. In~the scenario where the source is located near an observer, the~redshift can be approximated by a linear function of the distance $r$ when $z\ll 1$, consistent with observations. To~see this, we expand the expression in (\ref{eq:redshift}) as follows:
\begin{equation}\label{eq:redshift approximation}
z= H_0r+2H_0^2r^2+\cdots.
\end{equation}
This shows that the first term, $H_0r$, dominates the redshift at low values of $z$. In~fact, it is precisely this favorable result that motivated our selection of the function $\alpha_c(\lambda)$ as assumed in (\ref{eq:cosmic alpha}).

The close sources' redshift $z$ and distance $r$ exhibit a well-established and fundamental observational fact of a linear relationship, one that remains independent of cosmological models. Hence, it is imperative to subject any plausible cosmological model to rigorous testing using relevant observations. We validate our cosmological model by comparing the predicted luminosity distance $d_{\rm L}$with the values derived from SN Ia data of the MLCS2k2 Full Sample~\cite{Riess_2004}.

The luminosity distance is defined such that the Euclidean inverse-square law for the diminution in light with distance from a point source is preserved. Let $L$ denote the absolute luminosity of a source at distance $r$ and $l$ denote the observed apparent luminosity, and the~luminosity distance is defined as
\begin{equation}\label{eq:defining dL}
d_\text{L}=\left(\frac{L}{4\pi l}\right)^{1/2}.
\end{equation}
The area of a spherical surface centered on the source and passing through the observer is just $4\pi r^2$. The~photons emitted by the source arrive at this surface having been redshifted by the quantum effect by a factor $1+z$. We therefore find
\begin{equation}\label{eq:defing l}
l=\frac{L}{4\pi r^2}\frac{1}{1+z},
\end{equation}
from which
\begin{equation}\label{eq:dL and r}
d_\text{L}=r(1+z)^{1/2}.
\end{equation}
The coordinate distance $r$ can be expressed in terms of $z$ and $H_0$ from (\ref{eq:redshift}), we thus obtain
\begin{equation}\label{eq:dL and z}
d_\text{L}=\frac{\sqrt{1+z}}{H_0}\ln\left[\frac{1}{2}\left(3-\frac{1}{(1+z)^2}\right)\right].
\end{equation}
The distance modulus is defined by
\begin{equation}\label{eq:distance modulus}
\mu=m-M=5\log\left(\frac{d_\text{L}}{\text{Mpc}}\right)+25,
\end{equation}
where $m$ and $M$ are the apparent magnitude and absolute magnitude related to $l$ and $L$, respectively, and~we have explicitly specified that $d_\text{L}$ is measured in units of~megaparsecs.

\begin{figure}[H]
  \includegraphics[width=0.7\columnwidth]{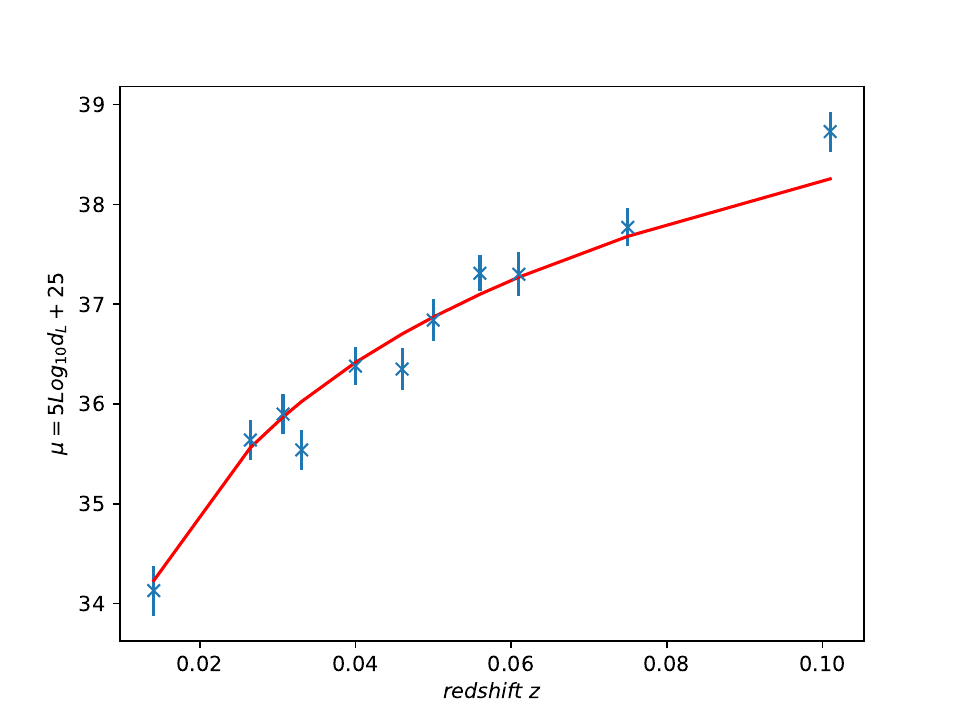}
  \caption{{The} 
 fitting results of the predicted distance moduli given by (\ref{eq:distance modulus}) to the data of the MLCS2k2 full sample~\cite{Riess_2004}. The~best fitted value of $H_0$ is $H_0=59.0_{-1.63}^{+1.68}$, and~$\chi^2=0.0018$.}\label{fig:redshift}
\end{figure}

We use the full cosmological sample of SNe Ia, MLCS2k2, presented in {Table~5} 
 of Riess~et~al.'s paper~\cite{Riess_2004}. The~sample consists of 11 SNe with redshift ranges from $0.01$ to $0.1$, which is ideal for testing our cosmological model. The~only free parameter in our model is the Hubble constant $H_0$, with~the best-fitted value being $H_0=59.0_{-1.63}^{+1.68}$ {km/s/Mpc}, as~presented in Figure~\ref{fig:redshift}.

Our cosmological model fits remarkably well with the SNe Ia data, providing evidence that the cosmological redshift may be due to QPE-induced time dilation rather than the expansion of the universe. Our model is based on a static universe, which is made possible by the assumption of quantum effects at cosmic scales, as described by QPE, acting as a repulsive force to balance gravity. When Einstein introduced the cosmological constant $\Lambda$ to balance gravity and obtain a static universe model, he failed, because $\Lambda$, regardless of its physical significance, is independent of the matter distribution. Consequently, any perturbations in matter result in unstable solutions of Einstein's equations. In~contrast, according to our proposal, QPE is not independent of matter distribution but is determined by the matter density, as~shown in Equation~(\ref{eq:rho_Q}). As~a result, any perturbations in matter distribution result in corresponding perturbations in QPE, to balance the extra gravity, allowing the universe to remain~static.

It is worth noting from (\ref{eq:rho_Q}) that QPE depends on two independent factors: $\rho^2_m(r)$ and $\alpha(\lambda)$. As~mentioned earlier, it is essential for  $\alpha(\lambda)$ to satisfy the condition that $\alpha(\lambda)$ approaches zero as $\lambda$ approaches zero. Thus, even though galaxies possess a high but finite central matter density, the~QPE would vanish in that region. However, in~the context of cosmology, the~QPE that is dependent on $\lambda$ displays some intricate yet justifiable characteristics. To~see this, let us rephrase the function $\alpha_c(\lambda)$ in (\ref{eq:cosmic alpha}) as follows:
\begin{equation}\label{eq:approximation cosmic alpha}
\alpha_c(\lambda)=\frac{2}{H_0\lambda}+\frac{1}{3}H_0\lambda+\cdots.
\end{equation}
It is evident that the first term $2/H_0\lambda$ plays a crucial role in generating a linear correlation between redshift and distance at a low redshift. At~first sight, the~inverse relationship between this term and $\alpha_c(\lambda)$ may appear to contradict the requirement that $\alpha(\lambda)$ approaches zero as $\lambda$ approaches zero. However, upon~further examination, this does not turn out to be the case. To~grasp this paradox, it is essential to recognize that it is $M_Q(\lambda)$, not $\rho_Q(r,\lambda)$, that determines the metric and thus the geometry of spacetime for a specific observer, as~shown in (\ref{eq:metric for cosmology}).

In cosmology, the~redshift of light signals emitted from distant sources is entirely attributed to the QPE, which varies with distance. However, this distance-dependent phenomenon does not imply any preferred direction in the universe, as~demonstrated both theoretically and~observationally.

{Although  we  prefer  the  presented  static  universe  model,  which  lacks  expansion,  a~ big  bang,  and~ evolution  of  the  entire  universe,  all  local  gravitational  systems,  such  as  galaxies  and  clusters  of  galaxies,  can  still  evolve.  In~ particular,  for~ macroscopic  systems,  such  as  the  solar  system,  planetary  systems,  and~ black  holes,  the~ evolution  will  continue  to  be  governed  by  general  relativity  as  usual. }

\subsection{Our Solution to the Galactic Rotation Curve~Problem}
After establishing the geometry of the entire universe's spacetime, we now shift our focus to examining the galaxies. Traditionally, based on Newtonian theory, when a stellar system achieves a state of equilibrium, the~kinetic energy supports the gravity, and~this equilibrium state must adhere to the virial theorem. However, astronomical observations indicate that the gravitational force exerted by ordinary matter is insufficient to counterbalance the kinetic energy of the system. As~a result, the~presence of a mysterious substance known as dark matter has been postulated to account for the discrepancy in mass, often referred to as the missing mass~problem.

Based on the theory presented in this paper, we can propose an alternative explanation that challenges the need for dark matter. In~fact, in~the mass expression provided in Equation~(\ref{eq:mass expression}), the~term QPE defined by Equation~(\ref{eq:rho_Q}) serves as the counterpart to dark~matter.

The mass expression presented in Equation~(\ref{eq:M and rhoQ}) is universally applicable to any radially symmetric system, making it suitable for investigating spherical galaxies. As~mentioned earlier, $\alpha(\lambda)$ should also be a universal function of scale $\lambda$. We assume a functional form given by (\ref{eq:functional form alpha}). In~the context of cosmology, the~scale can be regarded as arbitrarily large up to $r_h$, and the~reduced specific form of $\alpha_c(\lambda)$ provided in Equation~(\ref{eq:cosmic alpha}) has been demonstrated to align with observational data. For~galaxies, we use the parameters in (\ref{eq:functional form alpha}) with $A=0$ and $C=0$ to obtain
\begin{equation}\label{eq:alpha for galaxies}
\alpha_g(\lambda)=B(e^{H_0\lambda}-1),
\end{equation}
where, in~this paper, we consider $B$ as a constant, although~it may be a parameter, possibly scale-dependent, to~be determined by observations. When studying galaxies, the~scale $\lambda$ of interest is much smaller than the Hubble radius $1/H_0$, thus $\alpha_g(\lambda)$ can be approximated by $\alpha_g(\lambda)=BH_0\lambda$.

It is essential to account for the impact of the entire universe on local galaxies of finite mass, which is commonly referred to as the boundary condition when the metric of spacetime is concerned. Specifically, we mandate that the metric surrounding a local galaxy adheres to the condition that, as~$r$ approaches infinity, it converges to the metric of the entire universe (\ref{eq:metric for cosmology}), instead of the metric for flat spacetime, which has been conventionally used. To~achieve this, we replace $\rho_m(r)$ in Equation~(\ref{eq:M and rhoQ}) with $\rho_m(r)+\rho_c$. Since the impact of cosmological effects on local galaxies is unclear, we make the assumption that
\begin{equation}\label{eq:alpha_c for galaxies}
\alpha_c(\lambda)=\frac{A(\lambda)(e^{H_0\lambda}-1)}{H_0^2\lambda^2}-1
\end{equation}
when considering $\rho_c$, allowing us to determine the function $A(\lambda)$ based on observations of galaxies. We thus have from (\ref{eq:M and rhoQ})
\begin{equation}\label{eq:mass for galaxies}
\begin{split}
   M(r) =& 4 \pi\int_{0}^{r}(\rho_m(r')+\rho_c)r'^2dr'+4\pi\alpha_c(r)\int_{0}^{r}\rho_cr'^2dr'+\frac{4\pi\alpha_g(r)}{\rho_c}\int_{0}^{r}\rho_m^2(r')r'^2dr' \\
     =& M_b(r)+M_m(r)+M_Q(r),
\end{split}
\end{equation}
where
\begin{equation}\label{eq:three M}
\begin{split}
   M_b(r)=& \frac{A(r)}{2}(e^{H_0r}-1)r, \\
   M_m(r)=& 4\pi\int_{0}^{r}\rho_m(r')r'^2dr', \\
   M_Q(r)=& \frac{B4\pi(e^{H_0r}-1)}{\rho_c}\int_{0}^{r}\rho_m^2(r')r'^2dr' \\
         =& \frac{B32\pi^2(e^{H_0r}-1)}{3H_0^2}\int_{0}^{r}\rho_m^2(r')r'^2dr'.
\end{split}
\end{equation}
Evidently, $M_b(r)$ signifies the contribution of the quantum effects of the background mass density of the universe, $M_m(r)$ denotes the conventional mass, and $M_Q(r)$ represents the quantum effects of a galaxy itself. We thus have
\begin{equation}\label{eq:g_1 for galaxies}
\begin{split}
   g_1=& \left[1-\frac{2(M_b(r)+M_m(r)+M_Q(r))}{r}\right]^{1/2} \\
      =& \left[1-A(r)(e^{H_0r}-1)-\frac{2M_m(r)}{r}-\frac{2M_Q(r)}{r}\right]^{1/2}.
\end{split}
\end{equation}
We require $g_1$ for galaxies to satisfy the boundary condition, i.e.,~when $r\rightarrow\infty$, $g_1\rightarrow(3-2e^{H_0r})^{1/2}$, the~value for cosmology as shown in (\ref{eq:g1 for cosmology}). For~galaxies of finite mass or finite size, the~last two terms $\frac{2M_m(r)}{r}+\frac{2M_Q(r)}{r}\rightarrow 0$ when $r\rightarrow\infty$. We thus require $A(r)\rightarrow 2$ when $r\rightarrow\infty$. This condition imposes a nature constraint on the cosmological effect on local galaxies, as~shown~subsequently.

Our findings indicate that, from~the perspective of mass--energy, the~contributions of QPE to the metric exhibit no observable distinctions from conventional mass. Both QPE and conventional mass contribute to time dilation and distance contraction in precisely an identical manner, as~demonstrated in Equation~(\ref{eq:interval in spacetime}). As~such, it is reasonable to employ the conventional approach from general relativity when considering the stress--energy tensor and geometry of~spacetime.

Although our findings are presented in relativistic form, the~transition to a non-relativistic scenario for galaxies is straightforward. Actually, we can directly utilize the mass expression provided in Equation~(\ref{eq:mass for galaxies}) and apply Newtonian theory to investigate~galaxies.

Obviously, near~the center of a galaxy, the~gravitational effects arising from $M_b(r)$ and $M_Q(r)$ can be neglected when compared to those of $M_m(r)$. Towards the outer regions, both $M_b(r)$ and $M_Q(r)$ become significant, while the contribution of $M_m(r)$ diminishes with increasing radius $r$.

It is interesting to investigate the circular velocity of a test particle around a galaxy, which can be approximated by
\begin{equation}\label{eq:circular velocity}
\begin{split}
v^2(r)=&\frac{M(r)}{r} \\
   =&\frac{A(r)}{2} H_0r+\frac{4\pi}{r}\int_{0}^{r}\rho_m(r')r'^2dr'+\frac{B32\pi^2}{3H_0}\int_{0}^{r}\rho_m^2(r')r'^2dr',
\end{split}
\end{equation}
where we have neglected the terms with $H_0^nr^n$ when $n\geq 2$ for the scale of galaxies. For~a galaxy with a finite size $a$, when $r$ increases to the outer regions with $r\gg a$, we have
\begin{equation}\label{eq:approximated v(r)}
v(r)=\sqrt{\frac{A(r)}{2} H_0r+M_Q(a)/a}.
\end{equation}
This expression does not imply a constant value of $v(r)$ for large radii, as the~$H_0r$ term in (\ref{eq:approximated v(r)}) becomes more and more significant with increasing $r$. In~particular, this leads to a universal centrifugal acceleration (recall that $A(r)\rightarrow 2$ for $r\rightarrow\infty$)
\begin{equation}\label{eq:centrifugal acceleration}
\frac{v^2}{r}=cH_0,
\end{equation}
where $c$ is the speed of light. This result reflects the fact that the matter in the entire universe can have a local observable effect in galaxies, a~fact first discovered in MOND theory but that cannot be explained within the theory itself \citep{1983ApJ...270..365M,2015CaJPh..93..250M}. In~fact, our cosmological model yields the value of $H_0=59.0$, resulting {in} 
 $cH_0=5.74\times 10^{-8}~ \text{cm}/\text{sec}^2$. A~value which is close to, but~obviously larger, than that found for the universal acceleration parameter $a_0=1.2\times 10^{-8}~\text{cm}/\text{sec}^2$ of the MOND theory~\citep{McGaugh_2004,famaey:hal-02927744}. This fact has a natural explanation. As~an increasing function of $r$, $A(r)\rightarrow 2$ only at cosmological distances, typically hundreds of $\textsf{Mpc}$. The~value of $a_0$ is smaller than $cH_0$ only because it is detected at distances in the outer regions of galaxies, which are much smaller than the cosmological distance.  Notably, conformal gravity yields comparable results when fitting rotation curve data~\cite{MANNHEIM2006340,2019IJMPD..2844022M}. In~our theory, these phenomena are not considered mysteries. The~cosmological effects on local galaxies are the natural results of the boundary condition, which is determined by the requirement that the spacetime metric around an isolated galaxy, created by the mass (energy) specified in (\ref{eq:three M}), should align with that of the entire universe, as demonstrated in (\ref{eq:metric for cosmology}), when $r\rightarrow \infty$.

Despite these remarkable achievements, the~most effective way to validate our theory is by leveraging the abundant observational data on the rotation curves of flattened dwarf and spiral galaxies. As~our results were obtained for spherical systems, they must be converted into axisymmetric systems where cylindrical coordinates are more suitable. One could replicate the procedure for axisymmetric systems~\cite{Doran1996PRD,doran2003geometric}, similarly to what was performed for spherical systems in this paper. However, a~more efficient approach to achieve our objectives is to consider the relevant terms on the right-hand side of Equation~(\ref{eq:circular velocity}) as the gravitational potential generated by a point mass (or a mass element). Subsequently, we calculate the total potential for disk galaxies in cylindrical systems $(R,\phi,z)$~\citep{MANNHEIM2006340}.

The first term on the right-hand side of (\ref{eq:circular velocity}) is $A(r)cH_0r/2$. This is a linear potential originating from the quantum effect of the entire universe, and~thus is independent of specific local galaxies. We simply express the contribution of this term to the total circular velocity of a test particle on a thin-disk as
\begin{equation}\label{eq:cosmic velocity}
v_L^2(R)=A(R)cH_0R.
\end{equation}
In this paper, we initially overlook the gradually increasing nature of the function $A(R)$ and consider it as a constant $A_0$, which can be determined using rotation curve~data.

The second term on the right-hand side of (\ref{eq:circular velocity}) is the usual Newtonian potential. We write the corresponding potential in cylindrical coordinates as
\begin{equation}\label{eq:newtonian cylindrical potential}
\Phi_m(R,z)=-G\int_{0}^{\infty}dR'\int_{0}^{2\pi}d\phi'\int_{-\infty}^{\infty}dz'\frac{R'\rho_m(R',z')}{(R^2+R'^2-2RR'\cos\phi'+(z-z')^2)^{1/2}},
\end{equation}
where $G$ is the gravitational constant (we temporarily transition back from natural units from now on in this subsection). Inserting the cylindrical coordinate Green's function Bessel function expansion
\begin{equation}\label{eq:Green's function expansion}
\frac{1}{(R^2+R'^2-2RR'\cos\phi'+(z-z')^2)^{1/2}}=\sum_{-\infty}^{\infty}\int_{0}^{\infty}dkJ_m(kR)J_m(kR')e^{im(\phi-\phi')-k(|z-z'|)}
\end{equation}
into (\ref{eq:newtonian cylindrical potential}) yields
\begin{equation}\label{eq:newtonian potential 2}
\Phi_m(R,z)=-2\pi G\int_{0}^{\infty}dk\int_{0}^{\infty}dR'\int_{-\infty}^{\infty}dz'R'\rho_m(R',z')J_0(kR)J_0(kR')e^{-k|z-z'|}.
\end{equation}
For razor-thin exponential disks with $\rho_m(R,z)=\Sigma(R)\delta(z)=\Sigma_0e^{-R/R_0}\delta(z)$, where $\Sigma_0$ is the central surface mass density and $R_0$ is the disk scale length, the~potential is
\begin{equation}\label{eq:final form of newtonian potential}
\Phi_m(R)=-\pi G\Sigma_0R_0y[I_0(y/2)K_1(y/2)-I_1(y/2)K_0(y/2)],
\end{equation}
where $y\equiv\frac{R}{R_0}$, ~$I_n$ and $K_n$ are modified Bessel functions. If~we differentiate Equation~(\ref{eq:final form of newtonian potential}) with respect to $R$, we obtain the circular velocity contributed by the Newtonian potential
\begin{equation}\label{eq:circular velocity for newtonian potential}
v_m^2(y)=R\frac{\partial\Phi_m}{\partial R}=\frac{\pi c^2}{R_0}\left(\frac{GM_\odot}{c^2}\right)\left(\frac{\Sigma_0}{M_\odot}\right)R_0^2y^2[I_0(y/2)K_0(y/2)-I_1(y/2)K_1(y/2)].
\end{equation}

Another potential of the quantum effect for galaxies on the right-hand side of (\ref{eq:circular velocity}) is $(64\pi^2/3H_0)\int_{0}^{r}\rho_m^2(r')r'^2dr'$. This corresponds to a logarithmic potential, and~the circular velocity for the razor-thin exponential disk is
\begin{equation}\label{eq:quantum potential for galaxies}
\begin{split}
   v_Q^2(R)=&\frac{B8\pi G^2}{3cH_0R_0}\int_{0}^{R}dR'\int_{0}^{2\pi}d\phi'\int_{-\infty}^{\infty}dz'R'\Sigma^2(R')\delta(z') \\
            =&\frac{B4\pi^2G^2\Sigma_0^2R_0}{3cH_0}\left[1-\left(1+\frac{2R}{R_0}\right)e^{-2R/R_0}\right].
\end{split}
\end{equation}

The total circular velocity used to fit the data is therefore
\begin{equation}\label{eq:total velocity}
v(R)=\sqrt{v_L^2(R)+v_m^2(R)+v_Q^2(R)}.
\end{equation}

\begin{figure}[H]
  \includegraphics[width=0.8\columnwidth]{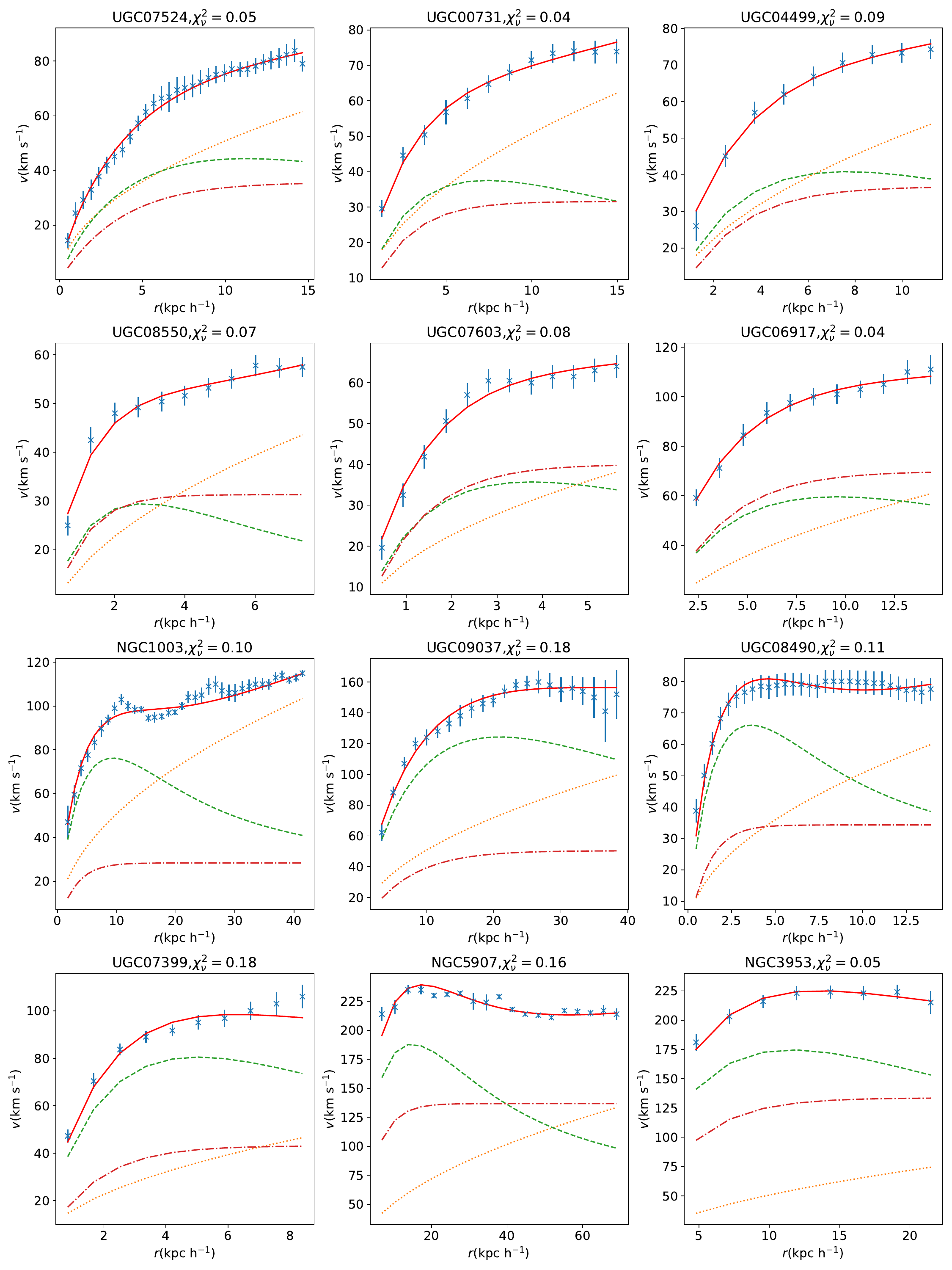}
  \caption{{The} 
 best-fitting rotation curves of LSB (upper 6 panels) and HSB (lower 6 panels) galaxies with the model presented in this paper. The~panels are listed by the increasing effective surface brightness of the galaxies. In~each panel, the~dotted curve shows the contribution from the cosmological quantum effect, given by (\ref{eq:cosmic velocity}); the dashed curve indicates the contribution from the luminous Newtonian potential given by (\ref{eq:circular velocity for newtonian potential}); the dash-dotted curve shows the contribution from the quantum effect of the galaxies themselves given by (\ref{eq:quantum potential for galaxies}); and finally, the~solid curve is the total circular velocity given by~(\ref{eq:total velocity}).}\label{fig:rotation curves}
\end{figure}

We fit the predicted circular velocity to the data provided in the Spitzer Photometry and Accurate Rotation Curves (SPARC) database~\cite{2016AJ....152..157L}. The SPARC database is the largest sample, to date, of rotation curves for every galaxy. It is a sample of 175 nearby galaxies with new surface photometry at 3.6 mu, and~high-quality rotation curves from previous HI/H studies. As~a first try, we consider each of the 175 galaxies in the sample as a razor-thin exponential disk characterized by a central surface mass density $\Sigma_0$ and a disk scale length $R_0$. In~addition to $\Sigma_0$ and $R_0$, we treat $A(r)=A_0$ in (\ref{eq:cosmic velocity}) and $B$ in (\ref{eq:quantum potential for galaxies}) as free parameters in our fitting. Figure~\ref{fig:rotation curves} displays 12 of them, including 6 low-surface-brightness (LSB) galaxies (6~upper panels) and 6 high-surface-brightness (HSB) galaxies (6 lower panels).

The fittings reveal that $A_0=0.02$ for all galaxies, with~$B=3.0$ for LSB galaxies and $B=0.3$ for HSB galaxies. The~cosmological effect on local galaxies thus suggests an acceleration of $A_0cH_0$$\sim$$10^{-9}~\text{cm}/\text{sec}^2$, significantly smaller than the universal value of $cH_0=5.74\times 10^{-8}~\text{cm}/\text{sec}^2$. Our fittings further suggest that $A_0$ can be substituted with
\begin{equation}\label{eq:A(R)}
A(R)=\frac{A_0+2(R/\text{Mpc})^2}{1+(R/\text{Mpc})^2}.
\end{equation}
This slowly varying function provides a good fit to the rotation curve data and also fulfills the requirement that $A(R)$ converges to $2$ as $R$ approaches cosmological distances. On~the other hand, the~value of B for LSB galaxies exceeds that for HSB galaxies, suggesting that LSB galaxies require relatively more QPE than HSB galaxies, and~$B(\lambda)$ is also scale~dependent.

All these results indicate that, in~terms of mass-energy, scale-dependent quantum phenomena do not exhibit a linear superposition, but~rather a hierarchical~process.

Despite the very simple model used to describe the matter distribution in LSB and HSB galaxies, the~fittings can still capture the main phenomenological results observed in other theories of gravity. For~instance, the~quantum effect contributions ($v_L^2+V_Q^2$) dominate most regions of LSB galaxies, whereas this is only true for the outer regions of HSB~galaxies.

Our proposed quantum effect on large scales has proven to be successful in explaining the mass discrepancy problem in~galaxies.

\section{Conclusions and~Discussions}\label{sec:conclusions}
We have proposed the quantum nature of spacetime at cosmic scales. To~investigate this postulation, we examined spherically symmetric and static  gravitational systems composed of free-falling particles with identical mass $m$. We assumed that the matter distribution can be smoothed using the {proper} number density $\rho(r)$ and {proper}  mass density $\rho_m(r)$ . We further assumed that the Dirac spinor $\psi(r)$ can fully capture all the physical aspects of the particles and satisfies the covariant Dirac equation. Therefore, we recognize $\rho(r)=\psi(r)\tilde{\psi(r)}$ and $\rho_m(r)=m\psi(r)\tilde{\psi(r)}=m\rho(r)$.  For~a given mass density $\rho_m(r)$, the~interplay between the number density $\rho(r)$ and the particle mass $m$ offers us a inherent mechanism to establish a connection between the spin density $S(r)=\frac{\hslash}{2}\psi(r)i\gamma_3\tilde{\psi(r)}$ and the scale $\lambda$ of gravitational~systems.

We utilized gauge theory gravity (GTG)~\cite{1998RSPTA.356..487L,1998JMP....39.3303D} to achieve the subsequent derivations. After~solving Einstein equations, the~metric of the spacetime for the radially symmetric and static gravitational systems was derived, as~shown in (\ref{eq:interval in spacetime}). The~fundamental results obtained thus far require additional physical interpretations and practical applications. We found that the quantum potential energy (QPE) defined by (\ref{eq:rho_Q})
\[
\rho_Q(r,\lambda)=-\frac{3}{4}\kappa S^2(r)=\rho_m^2(r)\alpha(\lambda)/\rho_c
\]
is contained in the metric via the mass (energy) $M(r)$, where $\alpha(\lambda)$ is a dimensionless function of the scale of the system involved.  The~mass within radius $r$ can then be expressed in (\ref{eq:M and rhoQ}).

It is important to note that at each radius $r$, we consider $M(r)$ as the mass of a gravitational system; therefore, we identify $r$ as the scale $\lambda$ of the ``system''.  As~such, for~any given mass density $\rho_m(r)$ of a gravitational system, we can derive the metric that encompasses not only the classical mass but also the quantum potential energy (QPE). When the scale of the gravitational systems are macroscopic, for~instance, the~solar system, the~QPE can be neglected, and~the Einstein--Newtonian theory of gravity is~recovered.

The dimensionless function $\alpha(\lambda)$ is universal, and {it should  be  determined  theoretically  from  first  principles.  However,  when  theoretical  formulas  are  lacking,  we  can  derive  a  phenomenological  formula  by  fitting  data  to  observations.}  We discovered that the general expression presented in (\ref{eq:functional form alpha}) successfully achieved our~objective.

While our fundamental results were initially derived for radially symmetric and static systems, they can be applied to any static gravitational system. By~substituting the proper mass density
$\rho_m$ with $\rho_m+\rho_Q++\rho_c+\rho_c\alpha_c(\lambda)$, where $\rho_c$ is the constant mass density of the universe, we establish a fundamental boundary condition that the spacetime metric encompassing any local gravitational system must adhere to. Remarkably, this boundary condition inherently gives rise to the cosmological impact on local galaxies, as~evidenced by observations of galactic rotation curves~\cite{famaey:hal-02927744,MANNHEIM2006340,McGaugh_2004,1983ApJ...270..365M,Milgrom:2014,2016AJ....152..157L}.

When applied to cosmology, our model yields a static universe. The~predicted luminosity distance--redshift relation fit remarkably well with SNe Ia data, providing evidence that the cosmological redshift may be due to QPE-induced time dilation rather than expansion of the universe. It is not surprising that when we extend Dirac's theory for free electrons to macroscopic particles (including stars within galaxies and galaxies within the entire universe), we are effectively moving away from the notion of a preferred absolute spacetime as suggested by general relativity and commonly accepted by many physicists, as~illustrated in the standard
$\Lambda$CDM cosmology. Without~the existence of absolute spacetime, physical processes should be described from the perspective of any observer in the universe independently and equivalently, without~relying on the absolute spacetime reference frame of the universe. Therefore, the~state of a celestial body should be characterized in the frame associated with a specific observer rather than in the frame of the absolute spacetime of the universe. Consequently, we assume that, for~any observer in the universe, a~distant celestial body behaves akin to a micro-particle and can be effectively described using Dirac's theory. On~the contrary, in~a scenario where absolute spacetime exists, all matter particles, including observers, must be described in relation to it. As~a result, the~motion of distant celestial bodies is typically explained by a combination of the local velocity and the overall expansion velocity. In~this context, we should not anticipate quantum effects at large scales for any~observers.

When considering galaxies, it is intriguing to note that quantum effects can serve as a substitute for dark matter. To~validate our theory, the~most effective approach was to utilize the extensive observational data on the rotation curves of dwarf and spiral galaxies. Remarkably, our proposed quantum effect at large scales successfully addressed the mass discrepancy issue in galaxies. Specifically, our theory presented a fitting formula, as outlined in (\ref{eq:A(R)}), which elegantly explains the correlation between the universal acceleration $cH_0$ suggested by cosmology, $a_0$ proposed by MOND, and~$0.02cH_0$  resulting from the cosmological effect on local galaxies~\cite{MANNHEIM2006340}.

{Although  not  investigated  in  the  present  study,  it  is  interesting  to  note  that  there  is  another  very  important  piece  of  evidence  supporting  our  new  theory  about  the  quantum  effects  of  macroscopic  matter  at  cosmic  scales.  This  evidence  comes  from  a  large-scale  structural  survey  of  the  universe,  which  reveals  the  fractal  geometry  of  matter  distribution at these scales,  as~ cited  in  Ref.~\cite{1998PhR...293...61S}.  It  has  now been  established  that  the  quantum  randomness  of  the  electron  can  be  mimicked  by  Brownian  motion;  the  simulated  trajectory  is  continuous  but  non-differentiable.  In~ fact,  the~ non-differentiable  trajectory  or  path  of  an  electron  exhibits  a  fractal  structure  due  to  the  uncertainty  principle  between  its  position  and  momentum  in  the  conventional  matrix  and  tensor  versions  of  quantum  mechanics~\cite{2000Fractal}.  Naturally,  the~ observations  of  the  fractal  geometry  of  the  large-scale  structure  of  the  universe  can  be  regarded  as  independent and  compelling  evidence  that  supports  our  assumption  when  extending  the  reasoning  about  the  relationship  between  the  quantum  nature  of  matter  and  its  fractal  path  from  microscopic matter  to  macroscopic  matter.}

In conclusion, we have expanded the quantum characteristics of microparticles to encompass macroscopic matter at cosmic scales. As~illustrated, this expansion was remarkably coherent. The~subsequent outcomes were derived within the framework of the firmly established theory of gravity (GTG) and mathematical language (STA). While there is room for improvement in the technical details of our derivations, the~fundamental discoveries are credible and, notably, were verified through astronomical observations of SNe Ia and galactic rotation curve data. In~particular, our proposal provides an alternative view point to understand the quantum nature of~spacetime.

The manifestation of physical laws should be independent of the mathematical language used, and~thus we believe that our proposal could also be accomplished, for~instance, through the tensor analysis approach. Of~course, using an unfamiliar language may make it challenging for readers to comprehend our theory. However, it has been shown that STA and GC can significantly streamline the calculations in our area of concern. Additionally, STA can illustrate the cohesive link between classical mechanics and quantum mechanics, thereby enhancing our comprehension of the enigmatic aspects of quantum theory and~spacetime.

{For  future  work,  we  could  test  our  theory  through  observations  of  gravitational  lensing,  cosmic  microwave  background  radiation,  and~ more  sophisticated  models  for  rotation  curves. In~particular, we could investigate the bullet cluster problem, which is considered the best evidence for the existence of dark matter.  A~ more  ambitious  project  would be  to  derive  the $\alpha(\lambda)$ formula from first principles.  }

\vspace{6pt}




\authorcontributions{{Conceptualization, D.-M.C.; methodology, D.-M.C.; software, L.W.; validation, D.-M.C.; formal analysis,D.-M.C.; investigation, D.-M.C.; resources, D.-M.C.; data curation, L.W.; writing---original draft preparation, D.-M.C.; writing---review and editing, D.-M.C.; visualization, L.W.; supervision, D.-M.C.; project administration, D.-M.C.; funding acquisition, D.-M.C.} 
}

\funding{{This work was supported by an NSFC grant (No. 11988101) and the K.C.Wong Education Foundation. } 
}

\dataavailability{{SPARC: http://astroweb.cwru.edu/SPARC/; SNe Ia data: Table 5 in Ref. [36] } 
}
\acknowledgments{We  would  like  to  express  our  appreciation  to  the  anonymous  reviewer  for  their  constructive  feedback  and  valuable  comments. 
}

\conflictsofinterest{{The authors declare no conflict of interest. } 
}


%
%
\appendixtitles{yes} 
\appendixstart
\appendix
\section[\appendixname~\thesection]{Spacetime Algebra and Geometric Calculus}\label{sec:STA}
 Spacetime algebra (STA) is generated by a 4-dimensional Minkowski vector space. The~inner and outer products of the four orthonormal basis vectors  \{$\gamma_{\mu}, \mu=0 \ldots 3$\} are defined to be
\begin{eqnarray}\label{eq:dot and wedge products for basis}
  \gamma_{\mu}\cdot\gamma_{\nu}\equiv\frac{1}{2}(\gamma_{\mu}\gamma_{\nu}+\gamma_{\nu}\gamma_{\mu}) &\equiv & \eta_{\mu\nu}=\text{diag}(+ - - -) \nonumber \\
  \gamma_{\mu}\wedge\gamma_{\nu} &\equiv &\frac{1}{2}(\gamma_{\mu}\gamma_{\nu}-\gamma_{\nu}\gamma_{\mu}).
\end{eqnarray}
A full basis for the STA is
\begin{equation}\label{eq:full set of STA}
1, \; \{\gamma_{\mu}\}, \; \{\sigma_{k}, i\sigma_{k}\}, \; \{i\gamma_{\mu}\}, \; i
\end{equation}
where $\sigma_k\equiv\gamma_k\gamma_0, k=1\ldots 3$ constitute the orthonormal basis of a 3-dimensional vector space relative to $\gamma_0$, and~$i=\gamma_0\gamma_1\gamma_2\gamma_3=\sigma_1\sigma_2\sigma_3$ is called pseudoscalar. The~STA is a linear space of dimensions $16$. We refer to the general elements of STA as multivectors, and~each multivector decomposes into a sum of elements of different grades. If~a multivector contains only grade-$r$ components, we call it homogeneous, and~is denoted by $A_r$. We call grade-$0$ multivectors scalars, grade-$1$ vectors, grade-$2$ bivectors and grade-$3$ trivectors. A~grade-$r$ multivector is called simple or an $r$-blade if and only if it can be written as
\begin{equation}\label{eq:blade}
A_r=a_1\wedge a_2\wedge ...\wedge a_r=\frac{1}{r!}\sum_{\pi}({\rm sgn} \pi)a_{\pi(1)}a_{\pi(2)}...a_{\pi(r)},
\end{equation}
where $\pi$ is a permutation of $1$ through $r$, `sgn $\pi$' is the sign of the permutation, $1$ for even and $-1$ for odd, and~the sum is over all $r!$ possible permutations. For~$r=2$, this reduces the outer product of two vectors
\begin{equation}\label{eq:outer product of two vectors}
a\wedge b=\frac{1}{2}(ab-ba),
\end{equation}
which is in agreement with the definition in (\ref{eq:dot and wedge products for basis}), and~clearly $a\wedge a=0$.

The geometric product of a grade-$r$ multivector $A_r$ with a grade-$s$ multivector $B_s$ is defined simply by $A_rB_s$, which decomposes into
\begin{equation}\label{eq:GA product}
A_rB_s=\langle A_rB_s\rangle_{|r-s|}+\langle A_rB_s\rangle_{|r-s|+2}+...+\langle A_rB_s\rangle_{r+s},
\end{equation}
where $\langle X\rangle_r$ denotes the projection onto the grade-$r$ part of $X$. The~grade-$0$ (scalar) part of $X$ is written as $\langle X\rangle$.
We employ ``$\cdot$'' and ``$\wedge$'' symbols to denote the lowest-grade and highest-grade terms in (\ref{eq:GA product}), so that
\begin{eqnarray}\label{eq:dot and wedge product}
  A_r\cdot B_s &=& \langle A_rB_s\rangle_{|r-s|},\label{eq:dot product Ar Bs} \\
  A_r\wedge B_s &=& \langle A_rB_s\rangle_{r+s},
\end{eqnarray}
which are called inner and outer products, respectively. They represent a generalization of the geometric product between two vectors $a$ and $b$, which can be expressed as follows:
\begin{equation}\label{eq:GA product ab}
ab=a\cdot b+a\wedge b.
\end{equation}

We define the reverse of a geometric product $AB$ by $(AB)^{\sim}=\tilde{B}\tilde{A}$, so that for vectors $a_1, a_2, \ldots, a_r$, we have
\begin{equation}\label{eq:reverse}
(a_1a_2...a_r)^{\sim}=a_ra_{r-1}...a_1.
\end{equation}
This reverse operator obeys the rule
\begin{eqnarray}
  (A+B)^{\sim} &=& \tilde{A}+\tilde{B}, \\
  \langle\tilde{A}\rangle &=& \langle A\rangle, \\
  \tilde{a} &=& a,
\end{eqnarray}
where $a$ is a vector. It is easy to show that
\begin{equation}\label{eq:reverse Ar}
 \tilde{A}_r=(-1)^{r(r-1)/2}A_r.
\end{equation}
Thus, suppose $r\leq s$, the~inner and outer products satisfy the symmetry properties
\begin{eqnarray}
  A_r\cdot B_s &=& (-1)^{r(s-1)}B_s\cdot A_r, \\
 A_r\wedge B_s &=& (-1)^{rs}B_s\wedge A_r.
\end{eqnarray}
The scalar product is defined by
\begin{equation}\label{eq:scalar product}
A*B=\langle AB\rangle.
\end{equation}
From (\ref{eq:dot product Ar Bs}), $A_r*B_s$ is nonzero only if $r=s$, thus the scalar product (\ref{eq:scalar product}) is commutative
\begin{equation}\label{eq:commutative scalar product}
\langle AB\rangle=\langle BA\rangle,
\end{equation}
which proves to be very useful in our calculations. The~inner product and outer product obey the distributive rule
\begin{eqnarray}
  A\cdot(B+C)=&=& A\cdot B+A\cdot C  \\
  A\wedge (B+C) &=& A\wedge B+A\wedge C.
\end{eqnarray}
The outer product is associative
\begin{equation}\label{eq:associative rule}
A\wedge (B\wedge C)=(A\wedge B)\wedge C.
\end{equation}
The inner product is not associative, but~homogeneous multivectors obey
\begin{eqnarray}
  A_r\cdot(B_s\cdot C_t) &=& (A_r\wedge B_s)\cdot C_t, \ \ {\rm for} \ \ r+s\leq t, \label{eq: inner product ABC} \\
  A_r\cdot(B_s\cdot C_t) &=& (A_r\cdot B_s)\cdot C_t, \ \ {\rm for} \ \ r+t\leq s.
\end{eqnarray}
For any vector $a$, it can be proved that
\begin{eqnarray}
  a\cdot A_r=\langle aA_r\rangle_{r-1} &=& \frac{1}{2}(aA_r-(-1)^rA_ra),  \\
  a\wedge A_r=\langle aA_r\rangle_{r+1} &=& \frac{1}{2}(aA_r+(-1)^rA_ra).
\end{eqnarray}
From which, we immediately have
\begin{equation}
aA_r=a\cdot A_r+a\wedge A_r.
\end{equation}
We also have the following useful {identities} 
\begin{subequations}
\begin{eqnarray}
  a\cdot(A_rB) &=& (a\cdot A_r)B+(-1)^rA_r(a\cdot B) \\
   &=& (a\wedge A_r)B-(-1)^rA_r(a\wedge B) \\
  a\wedge(A_rB) &=& (a\wedge A_r)B-(-1)^rA_r(a\cdot B) \\
   &=& (a\cdot A_r)B+(-1)^rA_r(a\wedge B).
\end{eqnarray}
\end{subequations}
These identities imply the following particularly useful relations:
\begin{eqnarray}
  a\cdot(A_r i) &=& (a\wedge A_r)i, \\
  a\wedge(A_r i) &=& (a\cdot A_r)i,
\end{eqnarray}
where $i$ is a pseudoscalar, and~we have used the fact that $a\wedge i=0$.

Any vector $a$ can be decomposed in terms of $\{\gamma_{\mu}\}$ into
\begin{equation}\label{eq:expanding a with frame}
a=a\cdot\gamma_{\mu}\gamma^{\mu}=a\cdot\gamma^{\mu}\gamma_{\mu},
\end{equation}
where the summation convention is implied. Similarly, an~arbitrary multivector $A$ can be decomposed into
\begin{equation}\label{eq:expanding multivector}
A=\sum_{\mu<...<\nu}A_{\mu...\nu}\gamma^{\mu}\wedge...\wedge\gamma^{\nu},
\end{equation}
where
\begin{equation}\label{eq:expanding coofficients for A}
A_{\mu...\nu}=A\cdot(\gamma_{\nu}\wedge...\wedge\gamma_{\mu}).
\end{equation}

We further define the commutator product
\begin{equation}\label{eq:commutator product}
A\times B=\frac{1}{2}(AB-BA),
\end{equation}
which satisfies the Jacobi identity
\begin{equation}\label{eq:Jocobi}
A\times(B\times C)+B\times(C\times A)+C\times(A\times B)=0.
\end{equation}
In addition, we have the Leibnitz rule for the commutator product
\begin{equation}\label{eq:Leibnitz rule for commutator product}
A\times(BC)=(A\times B)C+B(A\times C).
\end{equation}
This rule is particular useful with bivectors. It can be proved that
\begin{equation}\label{eq: commutator product with bivector preserves grade }
A_2\times A_r=\frac{1}{2}(A_2A_r-A_rA_2)=\langle A_2A_r\rangle_{r},
\end{equation}
which means that if one of the factors is a bivector, then the commutator product preserves its grade. Thus, for $\langle A\rangle_1=0$, we have
\begin{equation}
A_2A=A_2\cdot A+A_2\times A+A_2\wedge A.
\end{equation}
Since the commutator product with a bivector is grade preserving, the~identity in (\ref{eq:Leibnitz rule for commutator product}) still holds if $A=A_2$ and we replace all geometric products with either inner or outer products:
\begin{equation}\label{eq:Leibnitz rule for commutator product with bivectors}
\begin{array}{ccl}
A_2\times(B\cdot C)&=&(A_2\times B)\cdot C+B\cdot(A_2\times C), \\
A_2\times(B\wedge C)&=&(A_2\times B)\wedge C+B\wedge(A_2\times C).
\end{array}
\end{equation}

Finally, for~any vector $a$ and trivector $T$, it can be proved that
\begin{equation}\label{eq:commutator product of a and T}
(a\cdot T)\times T=0.
\end{equation}

Geometric calculus (GC) is the extension of a geometric algebra (like STA) to include differentiation and integration. Let the multivector $F$ be an arbitrary function of a multivector argument $X$, then the derivative of $F(X)$ with respective to $X$ in the $A$ direction is defined by
\begin{equation}\label{eq:multivector directional derivative}
A*\partial_{X}F(X)\equiv\lim_{\tau\rightarrow 0}\frac{F(X+\tau A)-F(X)}{\tau},
\end{equation}
where the multivector partial derivative $\partial_{X}$ inherits the multivector properties of its argument $X$. Suppose that the set $\{e_k\}$ form a vector frame (which is not necessarily orthonormal), the~reciprocal frame is determined by \citep{1986Hestenes+Sobczyk}
\begin{equation}\label{eq:reciprocal frame}
e^i=(-1)^{i-1}e_1\wedge e_2\wedge...\wedge \check{e}_i\wedge...\wedge_ne^{-1}
\end{equation}
\begin{equation}\label{eq:reciprocal pseudoscalar}
e\equiv e_1\wedge e_2\wedge...\wedge e_n
\end{equation}
and the check on $\check{e}$ denotes that this term is missing from the expression. As~usual, the~two frames are related by
\begin{equation}\label{eq:inner product of frame with reciprocal}
e_j\cdot e^k=\delta_j^k.
\end{equation}
From this frame, the multivector derivative $\partial_X$ in Equation~(\ref{eq:multivector directional derivative}) is defined by
\begin{equation}\label{eq:expanding of multivector derivative}
\partial_X\equiv\sum_{i<...<j}e^i\wedge...\wedge e^j(e^j\wedge...\wedge e_i)\ast\partial_X.
\end{equation}
The Leibnitz rule can be written in the form
\begin{equation}\label{eq:Leibnitz rule for multivector derivative}
\partial_X(AB)=\dot{\partial}_X\dot{A}B+\dot{\partial}_XA\dot{B},
\end{equation}
where the overdot indicates the scope of the multivector~derivative.

We have from (\ref{eq:multivector directional derivative}) and (\ref{eq:expanding of multivector derivative})
\begin{equation}\label{eq:partial derivative for scalar functions}
\partial_{X}\langle XA\rangle=P_{X}(A), \ \ \partial_X\langle\tilde{X}A\rangle=P_X(\tilde{A}),
\end{equation}
where $P_{X}(A)$ is the projection of $A$ onto the grades contained in $X$. These results are combined using Leibnitz's rule to give
\begin{equation}\label{eq:partial XX}
\partial_X\langle X\tilde{X}\rangle=\dot{\partial}_X\langle\dot{X}\tilde{X}\rangle + \dot{\partial}_X\langle X\dot{\tilde{X}}\rangle=2\tilde{X}.
\end{equation}

For a vector argument $x$ and a constant vector $a$, (\ref{eq:multivector directional derivative}) and (\ref{eq:partial derivative for scalar functions}) yield
\begin{equation}\label{eq:vector derivative for x}
a\cdot\partial_x x=a=\partial_x(x\cdot a).
\end{equation}

For a vector variable $a=a^{\mu}\gamma_{\mu}=a\cdot\gamma_{\mu}\gamma^{\mu}=a\cdot\gamma^{\mu}\gamma_{\mu}$, where $\gamma^{\mu}$ constitutes the reciprocal basis and satisfies $\gamma_{\mu}\cdot\gamma^{\nu}=\delta_{\mu}^{\nu}$, the~vector derivative can be defined as
\begin{equation}\label{eq:vector derivative a}
\partial_a\equiv\gamma^{\mu}\frac{\partial}{\partial a^{\mu}},
\end{equation}
For the derivative with respect to a spacetime position vector $x$, we use the symbol $\nabla\equiv\partial_x=\gamma^{\mu}\frac{\partial}{\partial x^{\mu}}$, if~$x=x^{\mu}\gamma_{\mu}$. From~(\ref{eq:vector derivative for x}) and (\ref{eq:vector derivative a}), we can obtain useful results
\begin{equation}\label{eq:vector partial derivative and basis}
\partial_a=\partial_{b}b\cdot\partial_a=\gamma^{\mu}\gamma_{\mu}\cdot\partial_a.
\end{equation}
Some results for the derivative with respect to position vector $x$ in an n-dimensional space are \citep{1986Hestenes+Sobczyk}
\begin{subequations}\label{eq: derivative aA}
\begin{eqnarray}
  \partial_x(x\cdot A_r) &=& rA_r, \label{eq:derivative xdotAr} \\
  \partial_x(x\wedge A_r) &=& (n-r)A_r, \\
  \dot{\partial}_xA_r\dot{x} &=& (-1)^{r}(n-2r)A_r.
\end{eqnarray}
\end{subequations}
From (\ref{eq: inner product ABC}) and (\ref{eq:derivative xdotAr}), an~r-blade $A_r$ can be expressed as
\begin{equation}\label{eq:express blade}
A_r=\frac{1}{r!}\partial_{a_1}\wedge ...\wedge\partial_{a_r}(a_r\wedge...\wedge a_1)\ast A_r.
\end{equation}

When considering a vector argument $x$ and a constant vector $a$, (\ref{eq:multivector directional derivative}) becomes the definition of the directional derivative $a\cdot\nabla$, thus
\begin{equation}\label{eq:directional derivative in the direction vector a}
a\cdot\nabla F=\left. a\cdot\partial_xF(x)=\frac{d}{d\tau}F(x+a\tau)\right|_{\tau=0}=\lim_{\tau\rightarrow 0}\frac{F(x+a\tau)-F(x)}{\tau}.
\end{equation}
Then, the general vector derivative can be obtained from the directional derivative using~(\ref{eq:vector partial derivative and basis}) as
\begin{equation}\label{eq:vector derivative from directional derivative}
\nabla F=\partial_xF(x)=\partial_a a\cdot\partial_xF(x).
\end{equation}

The directional derivative (\ref{eq:directional derivative in the direction vector a}) produces from $F$ a tensor field termed differential of $F$, denoted variously by
\begin{equation}\label{eq:differential of F}
\underline{f}(a)=F_a\equiv a\cdot\nabla F.
\end{equation}
The underbar notation serves to indicate that $\underline{f}(a)$ is a linear function of $a$. This induced linear function is very important for us to describe the apparatus of GC for handling transformations of spacetime and the induced transformations of multivector fields on~spacetime.

Suppose there is a diffeomorphism that transforms each point $x$ in some region of spacetime into another point $x'$ as
\begin{equation}\label{eq:mapping f}
x'=f(x).
\end{equation}
This induces a linear transformation of tangent vectors at $x$ to tangent vectors at $x'$ given by the differential
\begin{equation}\label{eq:induced transformation}
a'(x')=\underline{f}(a)=a\cdot\nabla f.
\end{equation}
If we regard $x$ as a map representing the ordering of points in spacetime, then $x'$ can be interpreted as a different map, or~a remapping of the same spacetime. The~transformation $f$ also induces an adjoint transformation $\bar{f}$, which takes a tangent vector $b'$ at $x'$ back to a tangent vector $b$ at $x$, as~defined by
\begin{equation}\label{eq:adjoint transformation}
b(x)=\bar{f}(b')\equiv\partial_xf(x)\cdot b'(x').
\end{equation}
The differential and its adjoint are related by
\begin{equation}\label{eq:relation between differential and its adjoint}
b'\cdot\underline{f}(a)=a\cdot\bar{f}(b').
\end{equation}
By using this relation, we can find one transformation from another by
\begin{eqnarray}\label{eq: find one transformation from another}
\bar{f}(a)&=&\partial_bb\cdot\bar{f}(a)=\partial_b(\underline{f}(b)\cdot a), \\
\underline{f}(a)&=&\partial_bb\cdot\underline{f}(a)=\partial_b(\bar{f}(b)\cdot a).
\end{eqnarray}

In addition to the induced linear transformations $\underline{f}(a)$ and $\bar{f}(a)$ of tangent vectors, by~the rule of direct substitution, (\ref{eq:mapping f}) can also induce a transformation of a multivector field $F(x)$ defined by
\begin{equation}\label{eq:the transformation of a F(x)}
F'(x')\equiv F'(f(x))=F(x),
\end{equation}
in which directional derivatives of the two functions are related by the chain rule
\begin{eqnarray}
  a\cdot\nabla F &=& a\cdot\partial_xF'(f(x)) \\
           &=& (a\cdot\nabla_xf(x))\cdot\partial_{x'}F'(x') \\
           &=& \underline{f}(a)\cdot\nabla'F'=a\cdot\bar{f}(\nabla')F'  \\
           &=& a'\cdot\nabla'F'.
\end{eqnarray}
The operator identity is
\begin{equation}\label{eq:operator identity for the chain rule}
a\cdot\nabla=\underline{f}(a)\cdot\nabla'=a\cdot\bar{f}(\nabla')=a'\cdot\nabla'.
\end{equation}
Differentiation with respect to the vector $a$ yields
\begin{equation}\label{eq:transformation law for nabla}
\nabla=\bar{f}(\nabla') \ \ {\rm or} \ \ \nabla'=\bar{f}^{-1}(\nabla).
\end{equation}

Now is an opportune moment to discuss linear algebra. In~fact, GC enables us to carry out coordinate-free calculations in linear algebra, eliminating the need for matrices. Every linear transformation $\underline{f}$ on spacetime has a unique extension to a linear function on the whole STA, called outermorphism. For~arbitrary multivectors $A$, $B$, and~any scalar $\alpha$, the~outermorphism is defined by the property
\begin{equation}\label{eq:outermorphism}
\underline{f}(A\wedge B)=\underline{f}(A)\wedge\underline{f}(B), \ \ \underline{f}(\alpha)=\alpha.
\end{equation}
It follows that for an $r$-blade $A_r=a_1\wedge...\wedge a_r$,
\begin{equation}\label{eq:outermorphism of r-blade}
\underline{f}(A_r)=\underline{f}(a_1)\wedge...\wedge\underline{f}(a_r).
\end{equation}
Since the outermorphism preserves the outer product, it also preserves grade:
\begin{equation}\label{eq:outermorphism preserves grade}
\underline{f}(\langle A\rangle_r)=\langle\underline{f}(A)\rangle_r
\end{equation}
for any multivector $A$. This implies that $\underline{f}$ alters the pseudoscalar $i$ only by a scalar multiple:
\begin{equation}\label{eq: f on pseudoscalar}
\underline{f}(i)=({\rm det}\underline{f})i,
\end{equation}
which defines the determinant of $\underline{f}$. The~product of two linear transformations $\underline{h}=\underline{g} \underline{f}$ also applies to their outermorphisms. It follows from (\ref{eq: f on pseudoscalar}) that
\begin{eqnarray}\label{eq:product of transformations g and f}
\underline{h}(i)=\underline{g}(\underline{f}(i))&=&(\det\underline{f}) \underline{g}(i)=(\det\underline{f})(\det\underline{g}) i \\
\det(\underline{g}\underline{f})&=&\det\underline{g}\det\underline{f}  \\
\det\underline{f}^{-1}&=&(\det\underline{f})^{-1}.
\end{eqnarray}
Every $\underline{f}$ has an adjoint $\bar{f}$, which can be extended to an outmorphism denoted by the same symbol
\begin{equation}\label{eq: adjoint to an outermorphism}
\langle A\bar{f}(B)\rangle=\langle B\underline{f}(A)\rangle
\end{equation}
for any multivectors $A$ and $B$. Unlike the outer product, the~inner product is not generally preserved by outermorphisms. However, it obeys the law
\begin{eqnarray}\label{eq:inner product of outermorphisms}
A_r\cdot\bar{f}(B_s)&=&\bar{f}\left(\underline{f}(A_r)\cdot B_s\right) \ \ {\rm for} \ \ r\leq s, \\
\underline{f}(A_r)\cdot B_s&=&\underline{f}\left(A_r\cdot\bar{f}(B_s)\right) \ \ {\rm for} \ \ r\geq s.
\end{eqnarray}
From these identities, we can construct a formula for the inverse of a linear transformation. Consider a multivector $B$, lying entirely in the algebra defined by the pseudoscalar $i$, we~have
\begin{equation}
\underline{f}(i)B=(\det\underline{f})i B=\underline{f}(i \bar{f}(B)).
\end{equation}
Replacing $i B$ by $A$, we find
\begin{equation}
(\det\underline{f})A=\underline{f}(i\bar{f}(i^{-1}A))
\end{equation}
with a similar result holding for the adjoint. It follows immediately that
\begin{eqnarray}
  \underline{f}^{-1}(A) &=& i\bar{f}(i^{-1}A)(\det\underline{f})^{-1}, \\
  \bar{f}^{-1}(A) &=& i\underline{f}(i^{-1}A)(\det\underline{f})^{-1}.
\end{eqnarray}

Once again, one great advantage of GC is that it eliminates unnecessary conceptual barriers between classical, quantum, and relativistic~physics.

\section[\appendixname~\thesection]{Stress--Energy Tensor Derived from Dirac Theory}\label{sec:ideal fluid}
In this Appendix, we show that the stress--energy tensor derived from the Dirac equation for free spin-$\frac{1}{2}$ particles  can be decomposed into the sum of a symmetric part and an antisymmetric part. The~symmetric part represents a classical pressureless ideal fluid, while the antisymmetric part represents the contribution of quantum potential energy.  As~a result, when spin is zero, the~stress--energy tensor becomes that of a pressureless ideal fluid. We follow Hestenes's work ~\citep{1973JMP....14..893H}, except~that we set $\beta=0$ for free spin-$\frac{1}{2}$ particles. As~mentioned in our previous paper, the~spinor field at spacetime point $x$ takes the form~\citep{Chen_2022}
\begin{equation}
  \psi(x)=\rho(x)^{1/2}R(x),
\end{equation}
where $\rho(x)$ is a scalar, representing the proper probability density and $R(x)$ is a rotor (Lorentz rotation) satisfying $R\tilde{R}=1$. The~rotor $R$ can be used to transform a fixed frame $\{\gamma_{\mu}\}$ into a new frame $\{e_{\mu}\}$
\begin{equation}\label{eq:new frame}
e_{\mu}=R\gamma_{\mu}\tilde{R}.
\end{equation}
We identify $v=e_0$ as the proper velocity associated with the expected history $x(\tau)$ of a particle, $v=\frac{dx}{d\tau}$. Our present objective is to derive a stress--energy tensor for a pressureless ideal fluid from the Dirac theory. The~desired form of the tensor is as follows:
\begin{equation}\label{eq:T for ideal fluid}
T_{\mu\nu}=\rho_m v_{\mu}v_{\nu},
\end{equation}
where $v=v_{\mu}\gamma^{\mu}$ and~$\rho_m=m\rho$, as explained in the previous~paper.

For later use, we define
\begin{eqnarray}
  \text{spin bivector}: S &=&\frac{1}{2}R\gamma_2\gamma_1\tilde{R}=\frac{1}{2}e_2e_1 \\
  \text{spin density trivector}: S_3 &=&\frac{1}{2}\psi i\gamma_3\tilde{\psi}=\frac{1}{2}\rho e_2e_1v=\rho S v. \label{eq:spin trivector}
\end{eqnarray}
The Dirac equation is
\begin{equation}\label{eq:Dirac equation}
\nabla\psi i\gamma_3=m\psi.
\end{equation}
A stress--energy tensor $T(a)$ is a linear vector function of a vector variable $a$, which denotes a flux of energy--momentum through a hypersurface with normal $a$ at spacetime point $x$. The~tensor $T(a)$ for the free Dirac field is given by~\citep{1973JMP....14..893H,doran2003geometric}
\begin{equation}\label{eq:stress-energy tensor}
T(a)=\gamma^{\nu}\langle a\partial_{\nu}\psi i\gamma_3\tilde{\psi}\rangle=\gamma^{\nu}a\cdot\langle\partial_{\nu}\psi i\gamma_3\tilde{\psi}\rangle_1.
\end{equation}
This tensor is not symmetric, and~its adjoint (transposed tensor) is
\begin{equation}\label{eq:adjoint of stress-energy tensor}
\bar{T}(a)=\partial_b\langle T(b)a\rangle=\langle a\cdot\nabla\psi i\gamma_3\tilde{\psi}\rangle_1.
\end{equation}
We can decompose $T(a)$ into a sum of a symmetric part and an anti-symmetric part as~\citep{1986Hestenes+Sobczyk}
\begin{equation}\label{eq:decompose T(a)}
T(a)=T_{S}(a)+T_{A}(a),
\end{equation}
where
\begin{equation}\label{eq:symmetric tensor}
T_{S}(a)=\frac{1}{2}(T(a)+\bar{T}(a))=\frac{1}{2}\partial_a(a\cdot T(a)),
\end{equation}
\begin{equation}\label{eq:anti-symmetric tensor}
T_{A}(a)=\frac{1}{2}(T(a)-\bar{T}(a))=\frac{1}{2}a\cdot(\partial_b\wedge T(b))=-\frac{1}{2}a\cdot(\partial_b\wedge\bar{T}(b)).
\end{equation}
From (\ref{eq:spin trivector})  and (\ref{eq:adjoint of stress-energy tensor}), we write
\begin{equation}\label{eq:anti-symmetric part of T}
\begin{split}
\partial_a\wedge\bar{T}(a)&=\dot{\nabla}\wedge\langle\dot{\psi} i\gamma_3\tilde{\psi}\rangle_1 \\
     &=\frac{1}{2}\langle\nabla\psi i\gamma_3\tilde{\psi}-\gamma^{\mu}(\psi i\gamma_3(\partial_{\mu}\psi)^{\sim})\rangle_2 \\
     &=-\frac{1}{2}\langle\nabla\psi i\gamma_3\tilde{\psi}+\gamma^{\mu}(\psi i\gamma_3(\partial_{\mu}\psi)^{\sim})\rangle_2 \\
     &=-\frac{1}{2}\langle\nabla(\psi i\gamma_3\tilde{\psi})\rangle_2 \\
     &=-\nabla\cdot S_3,
\end{split}
\end{equation}
where we have used the fact that the Dirac Equation~(\ref{eq:Dirac equation}) implies $\langle\nabla\psi i\gamma_3\tilde{\psi}\rangle_2=0$. Thus, according to (\ref{eq:anti-symmetric tensor}) and (\ref{eq:anti-symmetric part of T}), the~anti-symmetric part of $T(a)$ can be written as
\begin{equation}\label{eq:expression of TA}
T_A(a)=\frac{1}{2}a\cdot(\nabla\cdot S_3)=\frac{1}{2}(a\wedge\nabla)\cdot S_3.
\end{equation}

We are now ready to define the local energy--momentum $p$, one of the most fundamental quantities of the Dirac theory, as~\citep{1973JMP....14..893H}
\begin{equation}\label{eq:momentum}
\rho p=T(v)=v^{\mu}T_{\mu},
\end{equation}
where $v=v^{\mu}\gamma_{\mu}$ and $T_{\mu}=T(\gamma_{\mu})$ are understood. Now $T_{\mu}$ can be decomposed in the form~\citep{1973JMP....14..893H}
\begin{equation}\label{eq:decompose Tmu}
T_{\mu}=\rho v_{\mu}p+N_{\mu},
\end{equation}
where $N_{\mu}=N(\gamma_{\mu})$ describes the flow of energy momentum normal to the velocity. This can be verified from (\ref{eq:momentum}) and (\ref{eq:decompose Tmu}), which reads $N(v)=N(v^{\mu}\gamma_{\mu})=v^{\mu}N_{\mu}=0$, since $v^{\mu}v_{\mu}=v^2=1$.

According to the conservation of energy and momentum, the~divergence of a stress--energy tensor must be equal to the force density exerted on the particle. This relationship holds true even for free particles, where no external forces are acting on them. We thus write
\begin{equation}\label{eq:divergence of T}
\dot{T}(\dot{\nabla})=\dot{\bar{T}}(\dot{\nabla})=\dot{\bar{T}}(\gamma^{\mu}\dot{\partial_{\mu}})=\partial_{\mu}\bar{T}^{\mu}=0,
\end{equation}
where we have used the fact that the divergence of the anti-symmetric part of the stress--energy tensor vanishes. This is because, according to (\ref{eq:expression of TA}),  $\dot{T}_A(\dot{\nabla})=0$.

One advantage of STA is that the formulations of physical laws in Dirac theory, such as conservation laws and dynamics, are expressed in the same form as in classical mechanics. From~(\ref{eq:new frame}), it follows that
\begin{equation}\label{eq:differential of new frames}
a\cdot\nabla e_{\mu}=\Omega(a)\times e_{\mu},
\end{equation}
where
\begin{equation}\label{eq:angular velocity}
\Omega(a)=2a\cdot\nabla R\tilde{R}
\end{equation}
is a bivector valued function of a vector variable, which can be explained as the angular velocity of the frame $\{e_{\mu}\}$ rotates in the $a$ direction, and~$A\times B=\frac{1}{2}(AB-BA)$. From~(\ref{eq:stress-energy tensor}) and (\ref{eq:decompose Tmu}), we write
\begin{equation}\label{eq:Tmunu}
T_{\mu\nu}=T_{\mu}\cdot\gamma_{\nu}=\gamma_{\mu}\cdot\langle\partial_{\nu}\psi i\gamma_3\tilde{\psi}\rangle_1=\rho v_{\mu}p_{\nu}+N_{\mu\nu},
\end{equation}
where $p_{\mu}=p\cdot\gamma_{\mu}$ and $N_{\mu\nu}=N_{\mu}\cdot\gamma_{\nu}$. To~express $T_{\mu\nu}$ with observables, we need to calculate $\partial_{\nu}\psi i\gamma_3\tilde{\psi}$ from (\ref{eq:spinor}). We have
\begin{equation}\label{eq:vector1 in Tmunu}
\begin{split}
\partial_{\nu}\psi i\gamma_3\tilde{\psi}&=\partial_{\nu}(\rho^{1/2}R)i\gamma_3\rho^{1/2}\tilde{R} \\
     &=\frac{1}{2}\partial_{\nu}\rho R i\gamma_3\tilde{R}+\rho\partial_{\nu}R i\gamma_3\tilde{R} \\
     &=(\partial_{\nu}\ln\rho) S_3+\Omega_{\nu}S_3 \\
     &=(\partial_{\nu}\ln\rho)\rho S v+\rho\Omega_{\nu}Sv \\
     &=\rho(W_{\nu}-\partial_{\nu}S+\Omega_{\nu}S)v,
\end{split}
\end{equation}
Here, $\Omega_{\nu}=\Omega(\gamma_{\nu})=2\partial_{\nu}R\tilde{R}$; $W_{\nu}$ represents a bivector defined as
\begin{equation}\label{eq:definition of quantum potential}
W_{\nu}\equiv\frac{1}{\rho}\partial_{\nu}(\rho S),
\end{equation}
which is called quantum potential, and~this holds a pivotal position in quantum mechanics in the approach of causal interpretation~\citep{1973JMP....14..893H}. The~vector part of (\ref{eq:vector1 in Tmunu}) gives
\begin{equation}\label{eq:vector part of partial mu psi i gamma3tild psi}
\begin{split}
\langle\partial_{\nu}\psi i\gamma_3\tilde{\psi}\rangle_1 &=\Omega_{\nu}\cdot S_3 \\
     &=\rho(W_{\nu}\cdot v-\partial_{\nu}S\cdot v+\Omega_{\nu}\cdot S v+(\Omega_{\nu}\times S)\cdot v) \\
     &=\rho(W_{\nu}\cdot v+\Omega_{\nu}\cdot S v),
\end{split}
\end{equation}
where we have used $\partial_{\nu}S=\Omega_{\nu}\times S$, which can be proved as follows
\begin{equation}\label{eq:partial S}
\begin{split}
   \partial_{\mu}S &=\frac{1}{2}\partial_{\mu}(R\gamma_2\gamma_1\tilde{R}) \\
     &=\frac{1}{2}((\partial_{\mu}R)\gamma_2\gamma_1\tilde{R}+R\gamma_2\gamma_1\partial_{\mu}\tilde{R}) \\
     &=\frac{1}{2}(\Omega_{\mu}S-S\Omega_{\mu}) \\
     &=\Omega_{\mu}\times S.
\end{split}
\end{equation}
We thus obtain from (\ref{eq:stress-energy tensor}), (\ref{eq:momentum}) and (\ref{eq:vector part of partial mu psi i gamma3tild psi}),
\begin{equation}\label{eq:T(v)}
T(v)=v\cdot\langle\partial_{\nu}\psi i\gamma_3\tilde{\psi}\rangle_1\gamma^{\nu}=(v\wedge\Omega_{\nu})\cdot S_3\gamma^{\nu}=\rho\Omega_{\nu}\cdot S\gamma^{\nu}=\rho p,
\end{equation}
which means
\begin{equation}\label{eq:pnu}
p_{\nu}=\Omega_{\nu}\cdot S.
\end{equation}
Finally, from~(\ref{eq:Tmunu}), (\ref{eq:vector part of partial mu psi i gamma3tild psi}) and (\ref{eq:pnu}), we obtain~\citep{1973JMP....14..893H}
\begin{equation}\label{eq:Tmunu in terms observables}
T_{\mu\nu}=\rho v_{\mu}p_{\nu}+\rho(v\wedge\gamma_{\mu})\cdot W_{\nu}.
\end{equation}

To achieve the desired result (\ref{eq:T for ideal fluid}), we need to identify the constraints that would align the momentum and velocity (i.e., $v\wedge p=0$), and~ensure the symmetry of the stress--energy tensor (i.e., $T(a)=\bar{T}(a)$ or $T_{\mu\nu}=T_{\nu\mu}$).

Let us first multiply the Dirac Equation~(\ref{eq:Dirac equation}) on the right by $\tilde{\psi}$ to obtain
\begin{equation}\label{eq:Dirac eq multiply tilde psi}
\nabla\psi\gamma_2\gamma_1\tilde{\psi}=m\rho v.
\end{equation}
Next, we evaluate $\nabla\psi\gamma_2\gamma_1\tilde{\psi}$ by substituting $\psi$ from (\ref{eq:spinor}), resulting in
\begin{equation}\label{eq:left side of Dirac eq}
\begin{split}
   \gamma^{\mu}(\partial_{\mu}\psi\gamma_2\gamma_1\tilde{\psi}) &=\gamma^{\mu}((\partial_{\mu}\sqrt{\rho}) R\gamma_2\gamma_1\tilde{\psi}+\sqrt{\rho}(\partial_{\mu}R)\gamma_2\gamma_1\tilde{\psi}) \\
     &=\gamma^{\mu}((\partial_{\mu}\rho)S+\rho\Omega_{\mu}S) \\
     &=(\nabla\rho)S+\rho\gamma^{\mu}\Omega_{\mu}S.
\end{split}
\end{equation}
The vector part of this equation gives
\begin{equation}\label{eq:vector part of Dirac eq}
\begin{split}
\langle\nabla\psi\gamma_2\gamma_1\tilde{\psi}\rangle_1&= \langle(\nabla\rho)S+\rho\gamma^{\mu}\Omega_{\mu}S\rangle_1 \\
     &=(\nabla\rho)\cdot S+\rho\gamma^{\mu}\Omega_{\mu}\cdot S+\rho\gamma^{\mu}\cdot(\Omega_{\mu}\times S) \\
     &=(\nabla\rho)\cdot S+\rho p+\rho\nabla\cdot S \\
     &=\nabla\cdot(\rho S)+\rho p.
\end{split}
\end{equation}
From (\ref{eq:Dirac eq multiply tilde psi}) and (\ref{eq:vector part of Dirac eq}), we obtain
\begin{equation}\label{eq:momentum and velocity}
p=mv-\frac{1}{\rho}\nabla\cdot(\rho S).
\end{equation}

Decomposing the stress--energy tensor into its symmetric and anti-symmetric parts in terms of observables proves to be highly advantageous. In~the main text, we specifically utilize the adjoint form outlined in this Appendix, hence we express $\bar{T}_{\mu\nu}$ as derived from~(\ref{eq:Tmunu in terms observables}) as follows:
\begin{equation}\label{eq:decompose adjoint T}
\bar{T}_{\mu}=\rho p_{\mu}v+[\partial_{\mu}(S_3v)]\cdot v.
\end{equation}
By substituting $p$ given by (\ref{eq:momentum and velocity}), we have
\begin{equation}\label{eq:coordinate-free of adjoint of T}
\begin{split}
   \bar{T}(a)=& \rho a\cdot pv+[a\cdot\nabla(S_3v)]\cdot v \\
     =& \rho\left[ma\cdot v-\frac{1}{\rho}(a\wedge\nabla)\cdot(S_3v)\right]v+[a\cdot\nabla(S_3v)]\cdot v \\
     =&\rho_m a\cdot vv+[a\cdot\nabla(S_3v)]\cdot v-(a\wedge\nabla)\cdot(S_3v)v,
\end{split}
\end{equation}
where $\rho_m=\rho m$ represents the proper mass density. We see that if $S_3=0$, we promptly obtain the desired expression for a stress--energy tensor that is appropriate to describe a pressureless ideal fluid
\begin{equation}\label{eq:coordinate free form of energy momentum tensor}
T(a)=\bar{T}(a)=\rho_m a\cdot vv.
\end{equation}

\section[\appendixname~\thesection]{GTG and Matter}\label{sec:GTG}
By virtue of GC, GTG is constructed such that gravitational effects are described by a pair of gauge fields, $\bar{h}(a)=\bar{h}(a,x)$ and $\omega(a)=\omega(a,x)$, defined over a flat Minkowski background spacetime~\citep{1998RSPTA.356..487L}, where $x$ is the STA position vector, which is usually suppressed for~short.

The first of them, $\bar{h}(a)$, is a position-dependent linear function mapping the vector argument $a$ to vectors. The~introduction of $\bar{h}(a)$ ensures covariance of the equations under arbitrary local displacements (or an arbitrary remapping $x'=f(x)$) of the matter fields in the background spacetime. In~order to understand the physical meaning of the $\bar{h}$ field, we first define the covariant displacement transformation as
\begin{equation}\label{eq:covariant displacement}
M(x) \xrightarrow{x'=f(x)} M'(x)=M(x'),
\end{equation}
so that the equations $A(x)=B(x)$ and $A(x')=B(x')$ have exactly the same physical content. Suppose we have a vector field $b(x)=\nabla\phi(x)$, where $\phi(x)$ is a scalar field that is already covariant under displacement, i.e.,~$\phi'(x)=\phi(x')$. Now can we write $b'(x)=b(x')$ or $\nabla\phi'(x)=\nabla_{x'}\phi(x')$? Using the chain rule, we find
\begin{equation}\label{eq:chain rul}
\begin{array}{ccl}
  a\cdot\nabla\phi'(x) &=& a\cdot\nabla\phi(x')  \\
   &=& (a\cdot\nabla f(x))\cdot\nabla_{x'}\phi(x') \\
   &=& \underline{f}(a)\cdot\nabla_{x'}\phi(x') \\
   &=& a\cdot\bar{f}(\nabla_{x'})\phi(x'),
\end{array}
\end{equation}
where $\underline{f}(a)=a\cdot\nabla f(x)$ is a linear function of $a$ and an arbitrary function of $x$, and~$\bar{f}(\nabla_{x'})=\nabla f(x)\cdot\nabla_{x'}$, and~we call $\bar{f}$ the adjoint of $\underline{f}$, satisfying $a\cdot\underline{f}(b)=\bar{f}(a)\cdot b$, or~$\bar{f}(a)=\partial_b\langle\underline{f}(b)a\rangle$.
It follows that $\nabla\phi'(x)=\bar{f}(\nabla_{x'}\phi(x'))$, or~\begin{equation}\label{eq:nabla displacement}
\nabla_x=\bar{f}(\nabla_{x'}) \ \ \ \ \text{and} \ \ \ \ \bar{f}^{-1}(\nabla_x)=\nabla_{x'},
\end{equation}
which shows us that $b(x)$ is not covariant under displacement. In~order to make objects such as $b(x)$ covariant, we must introduce a position-gauge field $\bar{h}(a,x)$, which is a linear function of $a$ and arbitrary function of $x$, so that
\begin{equation}\label{eq:h field}
\bar{h}(a,x)\xrightarrow{x'=f(x)}\bar{h}'(a,x)=\bar{h}(\bar{f}^{-1}(a),x').
\end{equation}
Now, if we redefine $b(x)=\bar{h}(\nabla\phi(x))$, then
\begin{equation}\label{eq:covariant b}
\begin{array}{ccl}
b(x)=\bar{h}(\nabla\phi(x))\xrightarrow{x'=f(x)}b'(x)&=&\bar{h}'(\nabla\phi'(x)) \\
&=&\bar{h}(\bar{f}^{-1}(\nabla\phi'(x))) \\
&=&\bar{h}(\nabla_{x'}\phi(x'))=b(x'),
\end{array}
\end{equation}
which becomes covariant. The~$\bar{h}(a)$ field plays the same role as vierbein in the tensor calculus approach of gauge theory of gravity~\citep{1976RevModPhys.48.393,1998RSPTA.356..487L}. For~later use, we give the relationship between $\bar{h}(a)$ and the metric tensor $g_{\mu\nu}$ in GR. We define a position gauge invariant directional derivative as~\citep{2005FoPh...35..903H,1998RSPTA.356..487L}
\begin{equation}\label{eq:directional derivative}
L_a=a\cdot\bar{h}(\nabla)=\underline{h}(a)\cdot\nabla,
\end{equation}
where $\underline{h}$ is the adjoint of $\bar{h}$ defined by $\underline{h}(b)\equiv\partial_c(\bar{h}(c)\cdot b)$, and~$a$ is an invariant vector (i.e., for~$x'=f(x)$, $a(x)$ is transformed to $a'(x)=a(x')$). So, $\underline{h}$ maps tangent vectors to tangent vectors and $\bar{h}$ maps cotangent vectors to cotangent vectors.  For~a given set of coordinates \{$x^{\mu}, \mu=0, 1, 2, 3$\}, we introduce the basis vectors
\begin{equation}\label{eq:basis}
e_{\mu}\equiv\frac{\partial x}{\partial x^{\mu}}, \ \ \ e^{\mu}\equiv\nabla x^{\mu},
\end{equation}
which satisfy $e_{\mu}\cdot e^{\nu}=\delta^{\nu}_{\mu}$. From~these vectors, we further define vectors
\begin{equation}\label{eq:square root of metric}
g_{\mu}\equiv\underline{h}^{-1}(e_{\mu}), \ \ \ g^{\mu}\equiv\bar{h}(e^{\mu}).
\end{equation}
These vectors satisfy the relation
\begin{equation}\label{eq:orthonormal basis}
g_{\mu}\cdot g^{\nu}=\underline{h}^{-1}(e_{\mu})\cdot\bar{h}(e^{\nu})=e_{\mu}\cdot\bar{h}^{-1}(\bar{h}(e^{\nu}))=e_{\mu}\cdot e^{\nu}=\delta^{\nu}_{\mu}.
\end{equation}
The metric tensor is then given by
\begin{equation}\label{eq:metric}
  g_{\mu\nu}\equiv g_{\mu}\cdot g_{\nu}.
\end{equation}

Let $x(\tau)$ be a time-like curve (where $\tau$ is the proper time), a~mapping $f: x\rightarrow x'=f(x)$ induces the transformation
\begin{equation}\label{eq:map velocity}
\dot{x}=\frac{d x}{d\tau}\rightarrow \dot{x}'=\frac{d x'}{d\tau}=\frac{d x}{d\tau}\cdot\nabla f(x).
\end{equation}
Comparing Equation~(\ref{eq:map velocity}) with Equation~(\ref{eq:directional derivative}), we introduce an invariant velocity $v=v(x(\tau))$ as~\citep{2005FoPh...35..903H}
\begin{equation}\label{eq:invariant velocity}
\dot{x}=\underline{h}(v), \ \ \ v=\underline{h}^{-1}(\dot{x}).
\end{equation}
From the known formula $dx=dx^{\mu}e_{\mu}$, the~invariant normalization $v^2=1$ induces the invariant line element on a time-like curve in GR
\begin{equation}\label{eq:line element}
d\tau^2=[\underline{h}^{-1}(dx)]^2=g_{\mu\nu}dx^{\mu}dx^{\nu}.
\end{equation}

Another gauge field, $\omega(a)$, is a position-dependent linear function mapping the vector argument $a$ to bivectors. Its introduction ensures covariance of the equations of physics under local Lorentz rotations described by the rotor $R$.  Under~local Lorentz rotations, the~multivector $M$ transforms as $M'=RM\tilde{R}$ and the spinor $\psi$ transforms as $\psi'=R\psi$. To~ensure covariance of the quantities like $\bar{h}(\nabla)M$ and $\bar{h}(\nabla)\psi$ under local Lorentz rotations, $\bar{h}(\nabla)=\bar{h}(\partial_a)a\cdot\nabla$ should be replaced by a covariant derivative $D$ ~\citep{1998RSPTA.356..487L}. To~achieve this, we focus attention on $a\cdot\nabla\psi'=a\cdot\nabla(R\psi)$ and write
\begin{equation}\label{eq:directional derivative on psi}
a\cdot\nabla(R\psi)=Ra\cdot\nabla\psi+(a\cdot\nabla R)\psi.
\end{equation}
Clearly, the~presence of the term $(a\cdot\nabla R)\psi$ renders the operator $a\cdot\nabla$ non-covariant. Since the rotor $R$ satisfies $R\tilde{R}=1$, we find that
\begin{equation}\label{eq:RtildeR=1}
a\cdot\nabla R\tilde{R}+R a\cdot\nabla\tilde{R}=0,
\end{equation}
which implies
\begin{equation}\label{eq:a cdot nabla R tilde R}
a\cdot\nabla R\tilde{R}=-R a\cdot\nabla\tilde{R}=-(a\cdot\nabla R\tilde{R})^{\sim}.
\end{equation}
Hence, $a\cdot\nabla R\tilde{R}$ is equal to minus its reverse and thus must be a bivector in the STA. We can therefore rewrite (\ref{eq:directional derivative on psi}) as
\begin{equation}\label{eq: alternative form for the derectional derivative on psi }
a\cdot\nabla(R\psi)=R a\cdot\nabla\psi+\frac{1}{2}(2a\cdot\nabla R\tilde{R})(R\psi).
\end{equation}
This suggests that to~achieve a covariant derivative we must add a connection term to $a\cdot\nabla$ to construct an operator
\begin{equation}\label{eq:Da psi}
D_a\psi=a\cdot\nabla\psi+\frac{1}{2}\Omega(a)\psi
\end{equation}
which must be covariant under local Lorentz rotations. The~connection $\Omega(a)=\Omega(a,x)$ is a bivector valued linear function of $a$ with an arbitrary $x$ dependence. Under~local rotations we expect that the operator $D_a$ will be unchanged in form, namely,
\begin{equation}\label{eq:Da prime}
D_a'=a\cdot\nabla+\frac{1}{2}\Omega'(a),
\end{equation}
Here, we have used the fact that $a\cdot\nabla$ cannot change under local rotations. Nevertheless, the~property that the covariant derivative must satisfy is
\begin{equation}\label{eq:Da Prime psi prime}
\begin{array}{ccl}
D_a'\psi'&=&(a\cdot\nabla+\frac{1}{2}\Omega'(a))(R\psi) \\
&=&RD_a\psi=R(a\cdot\nabla\psi+\frac{1}{2}\Omega(a)\psi).
\end{array}
\end{equation}
From these identities, it follows that $\Omega(a)$ transforms as
\begin{equation}\label{eq:Omega to Omega'}
\Omega'(a)=R\Omega(a)\tilde{R}-2a\cdot\nabla R\tilde{R}.
\end{equation}
Now, we reassemble the covariant derivative (\ref{eq:Da psi}) with the term $\bar{h}(\partial_a)$ to form
\begin{equation}\label{eq:D}
D\equiv\bar{h}(\partial_a)D_a,
\end{equation}
and write
\begin{equation}\label{eq:D psi}
D\psi=\bar{h}(\partial_a)(a\cdot\nabla\psi+\frac{1}{2}\Omega(a)\psi).
\end{equation}

Note that the vector derivative $D$ is fully covariant, but~$D_a$ is not, since it contains the $\Omega(a)$ field, which must transform in the same way as $a\cdot\nabla R\tilde{R}$ under displacement and thus picks up a term in $\underline{f}$ (\ref{eq:operator identity for the chain rule}). Recall the definition of the position gauge invariant directional derivative (\ref{eq:directional derivative}), we thus define
\begin{equation}\label{eq:a.D}
a\cdot D\psi=a\cdot\bar{h}(\nabla)\psi+\frac{1}{2}\omega(a)\psi,
\end{equation}
where $\omega(a)$ is defined by
\begin{equation}\label{eq:omega and Omega}
\omega(a)=\Omega(h(a)).
\end{equation}
It should be pointed out that the vector $a$ is declared to be position gauge-invariant, as~stated below Equation~(\ref{eq:directional derivative}). Therefore, $D_a$ is related to $a\cdot D$ by
\begin{equation}\label{eq:Da psi and a.D psi}
\begin{array}{ccl}
D_a\psi&=&\underline{h}^{-1}(a)\cdot D\psi \\
&=&\underline{h}^{-1}(a)\cdot\bar{h}(\nabla)\psi+\frac{1}{2}\omega(\underline{h}^{-1}(a))\psi \\
&=&a\cdot\nabla\psi+\frac{1}{2}\Omega(a)\psi,
\end{array}
\end{equation}
or simply
\begin{equation}\label{eq:Da and a.D}
D_a=\underline{h}^{-1}(a)\cdot D.
\end{equation}
We thus also have
\begin{equation}\label{eq:covariant derivative on psi}
D\psi\equiv\partial_a\left(a\cdot\bar{h}(\nabla)\psi+\frac{1}{2}\omega(a)\psi\right).
\end{equation}

We are now ready to establish the form of the covariant derivatives of the observables formed from a spinor field. In~general, such observables have the form
\begin{equation}\label{eq:observables}
M=\psi\Gamma\tilde{\psi},
\end{equation}
where $\Gamma$ is a constant multivector formed from combinations of $\gamma_0$, $\gamma_3$ and $i\sigma_3$. Under~displacements $x'=f(x)$, the~spinor fields are covariant: $\psi'(x)=\psi(x')$. Thus, the observable $M$ inherits its transformation properties from the spinor $\psi$,
\begin{equation}\label{eq:displacement M}
M(x)\xrightarrow{x'=f(x)} M'(x)=M(x'),
\end{equation}
exactly the same form in (\ref{eq:covariant displacement}). Under~rotations, the~spinor transforms as $\psi'=R\psi$. So it follows from (\ref{eq:observables}) that, under~rotations, $M$ transforms as
\begin{equation}\label{eq:rotations M}
M(x)\xrightarrow{R} M'=\psi'\Gamma\tilde{\psi'}=R\psi\Gamma\tilde{\psi}\tilde{R}=RM\tilde{R}.
\end{equation}
To achieve the fully covariant derivative for $M$, we first write
\begin{equation}\label{eq:La M}
L_a M=L_a\psi\Gamma\tilde{\psi}+\psi\Gamma (L_a\psi)^{\sim}.
\end{equation}
This equation is a displacement covariant. We simply replace $L_a$ with $a\cdot D$ to obtain
\begin{equation}\label{eq:a.DM}
\begin{array}{ccl}
a\cdot DM&=&a\cdot D\psi\Gamma\tilde{\psi}+\psi\Gamma(a\cdot D\psi)^{\sim} \\
&=&L_a(\psi\Gamma\tilde{\psi})+\omega(a)\times(\psi\Gamma\tilde{\psi}) \\
&=&L_a M+\omega(a)\times M.
\end{array}
\end{equation}
Note the difference between the forms of $a\cdot D$ acting on $M$ and $\psi$.

The field strength corresponding to the $\omega(a)$ gauge field is defined by
\begin{equation}\label{eq:field strength for omega}
[a\cdot D, b\cdot D]\psi=\frac{1}{2}\mathcal{R} (a\wedge b)\psi,
\end{equation}
where
\begin{equation}\label{eq:curvature}
\mathcal{R}(a\wedge b)\equiv L_a\omega(b)-L_b\omega(a)+\omega(a)\times\omega(b),
\end{equation}
$a$ and $b$ are constant vectors. The Ricci tensor $\mathcal{R}(a)$, Ricci scalar $\mathcal{R}$, and Einstein tensor $\mathcal{G}(a)$ are defined, respectively, as
\begin{eqnarray}
  \mathcal{R}(a) &=& \partial_b\cdot\mathcal{R}(b\wedge a), \\
  \mathcal{R} &=& \partial_a\cdot \mathcal{R}(a), \\
  \mathcal{G}(a) &=& \mathcal{R}(a)-\frac{1}{2}a\mathcal{R}.
\end{eqnarray}

The overall action integral is of the form
\begin{equation}\label{eq:action}
I=\int|d^4x|\det(\textsf{h})^{-1}(\frac{1}{2}\mathcal{R}-\kappa\mathcal{L}_m),
\end{equation}
where $\mathcal{L}_m$ describes the matter content and $\kappa=8\pi$. In~this paper, we adopt the covariant  $\mathcal{L}_m=D i\gamma_3\tilde{\psi}-m\psi\tilde{\psi}$ from Dirac theory to describe  the macroscopic matter, which has been well-studied \citep{1998RSPTA.356..487L,1998JMP....39.3303D,1997GReGr..29.1527C} for electrons. From~(\ref{eq:action}), we obtain the following equations that describe the field coupled self-consistently to gravity~\citep{1998JMP....39.3303D}:
\begin{eqnarray}
  \text{torsion:}\quad   D\wedge\bar{h}(a) &=& \kappa \bar{h}(a)\cdot S, \label{eq:torsion} \\
  \text{Einstein:}\qquad  \mathcal{G}(a) &=& \kappa \mathcal{T}(a) \label{eq:Einstein equation} \\
  \text{Dirac:}\qquad    D\psi i\sigma_3 &=& m\psi\gamma_0, \label{eq:covariant Dirac equation}
\end{eqnarray}
where $D\wedge\bar{h}(a)$ is the gravitational torsion which is determined by the trivector spin density $S\equiv\frac{1}{2}\psi i\gamma\tilde{\psi}$, $\kappa=8\pi$, and~\begin{equation}\label{eq:energy-momentum tensor}
\mathcal{T}(a)=\langle a\cdot D\psi i\gamma_3\tilde{\psi}\rangle_1
\end{equation}
is the matter stress--energy tensor. We can solve Equation~(\ref{eq:torsion}) for $\omega(a)$ to obtain~\citep{1998JMP....39.3303D}
\begin{equation}\label{eq:omega}
  \omega(a)=\omega'(a)+\frac{1}{2}\kappa a\cdot S=-H(a)+\frac{1}{2}a\cdot[\partial_b\wedge H(b)]+\frac{1}{2}\kappa a\cdot S,
\end{equation}
this defines $\omega'(a)$ as the $\omega$-function in the absence of torsion, and~\begin{equation}\label{eq:H function}
H(a)\equiv \bar{h}(\dot{\nabla}\wedge\dot{\bar{h}}^{-1}(a))=-\bar{h}(\dot{\nabla})\wedge\dot{\bar{h}}(\bar{h}^{-1}(a)),
\end{equation}
where `overdot' notation is employed to denote the scope of a differential operator. This proves convenient if we employ the primed symbols to denote the torsion-free part of the curvature tensors, and we obtain~\citep{1998JMP....39.3303D}
\begin{eqnarray}
  \mathcal{R}(a\wedge b) &=& \mathcal{R}'(a\wedge b)+\frac{1}{4}\kappa^2[(a\wedge b)\cdot S]\cdot S \nonumber \\
               & &-\frac{1}{2}\kappa[(a\wedge b)\cdot D]\cdot S, \\
  \mathcal{R}(a) &=& \mathcal{R}'(a)+\frac{1}{2}\kappa^2(a\cdot S)\cdot S \nonumber \\
       & &-\frac{1}{2}\kappa a\cdot(D\cdot S), \\
  \mathcal{R} &=& \mathcal{R}'+\frac{3}{2}\kappa^2S^2.
\end{eqnarray}

%

\begin{adjustwidth}{-\extralength}{0cm}

\reftitle{References}

\PublishersNote{}
\end{adjustwidth}
\end{document}